\UseRawInputEncoding
\documentclass[11pt,aps,amssymb,prd,a4paper,nofootinbib]{revtex4-1}
\usepackage{fullpage}
\usepackage{amsfonts}
\usepackage{amsmath}
\usepackage{slashed}
\usepackage{amssymb}
\usepackage{graphicx}
\usepackage{makeidx}
\usepackage{cancel}
\usepackage{epic}
\usepackage{eepic}
\usepackage{epsfig}
\usepackage{latexsym}
\usepackage[dvipsnames]{xcolor}
\usepackage{float}
\usepackage{multirow}
\usepackage[export]{adjustbox}
\usepackage{xurl,hyperref}
\usepackage{enumitem}
\hypersetup{colorlinks=true,citecolor=red,linkcolor=NavyBlue,urlcolor=NavyBlue}
\usepackage[utf8]{inputenc}
\usepackage[caption=false]{subfig}
 
\usepackage{natbib}
\usepackage{relsize}
\usepackage[left=2.2 cm,right=2.2 cm,top=2 cm,bottom=2 cm]{geometry}
\usepackage{mathptmx}
\begin{document}
\relscale{1.05}
\captionsetup[subfigure]{labelformat=empty}

\title{Exploring the Null Results in the Direct Detection Experiments, $(g-2)_\ell$ and Neutrino Mass in an Extended $U(1)_{L_\mu-L_\tau}$ Model Constrained through the $Z\to\ell^+\ell^-$ Decays}
 
\author{Bibhabasu De}
\email{bibhabasude@gmail.com}
\affiliation{Department of Physics, The ICFAI University Tripura, Kamalghat-799210, India}

\date{\today}

\begin{abstract}
\noindent
The Direct Detection~(DD) experiments are vital for probing the particle nature of Dark Matter~(DM). However, in the absence of a scattering event, DD searches result in stringent bounds on the corresponding parameter space. The paper has considered a $U(1)_{L_\mu-L_\tau}$-extension of the Standard Model~(SM) and augmented the particle spectrum with $SU(2)_L$-singlet vector-like leptons and scalars. A discrete $Z_2$ symmetry stabilizes the lightest SM-singlet vector-like lepton as the viable DM candidate. In the proposed model, amplitude-level cancellation can be achieved for both DM-electron and DM-quark scatterings, leading to a trivial explanation for the continuous null results in the DD experiments. The framework can also induce one-loop corrections to the lepton anomalous magnetic moments and $Z\ell^+\ell^-$ couplings. The experimental bounds on the $Z\to\ell^+\ell^-$ decays are instrumental in constraining the model parameters. Particularly, using the $Z\to\tau^+\tau^-$ decay, a stronger exclusion limit can be imposed on the $U(1)_{L_\mu-L_\tau}$ parameter space. Further, in the presence of three heavy right-handed neutrinos, transforming as $Z_2$-even states, the model can explain all the neutrino mass and mixing constraints using the Type-I seesaw mechanism. Future experimental updates on the $(g-2)_\ell$, $Z\to\ell^+\ell^-$ decays and improved bounds on the $U(1)_{L_\mu-L_\tau}$ theory can be crucial to test the proposed model. Moreover, future DD experiments searching for a DM-muon scattering might be significant to probe the considered DM-SM interaction.
\end{abstract}
	
\maketitle	

\section{Introduction}
\noindent
The Standard Model has already been tested to a remarkable level of precision through various experiments, including the high-energy collider searches. The discovery of the 125 GeV SM-like scalar at the Large Hadron Collider~(LHC)~\cite{ATLAS:2012yve, CMS:2012qbp}, and the remeasurement of the $W$-boson mass at the ATLAS~\cite{ATLAS:2024erm} and CMS~\cite{CMS:2024lrd} can be mentioned as a few recent experimental outcomes that substantiate the robustness of the SM. Further, the 2025 update on the SM-prediction for the muon anomalous magnetic moment~\cite{Aliberti:2025beg} has significantly consolidated its acceptance within the current experimental precision. However, despite all its success as a highly predictive gauge formulation for the electroweak~(EW) and strong interactions, SM fails to explain several experimental, astrophysical, and cosmological observations --- the most intriguing one being the Dark Matter~(for a comprehensive review, see Refs.~\cite{GONDOLO1991145, Jungman:1995df, Bertone:2004pz,Cirelli:2024ssz}). Though the existence of the DM has been well-established through its gravitational signatures~\cite{1932BAN.....6..249O,Zwicky:1933gu, Zwicky:1937zza,Metcalf:2003sz,Planck:2018vyg}, to theorize the particle nature of the DM, one must extend the standard model. Weakly interacting massive particles~(WIMPs) represent a significant class of viable DM candidates as they require a minimal augmentation of the SM particle spectrum and have been exhaustively explored in the literature~\cite{Jungman:1995df, Roszkowski:2017nbc,Schumann:2019eaa,Smirnov:2019ngs}. Though such Beyond the Standard Model~(BSM) theories with a WIMP can easily explain the observed DM abundance~\cite{Planck:2018vyg}, are either under severe threat or have been ruled out through the DD experiments~\cite{PandaX-4T:2021bab, LZ:2022lsv, XENON:2023cxc, PandaX:2024qfu}~(for a recent review, see Ref.~\cite{Arcadi:2024ukq}). Naturally, the focus has been shifted to the sub-GeV mass regime where the DD bounds are slightly relaxed~\cite{XENON:2019gfn, XENON:2019zpr, SENSEI:2020dpa, PandaX-II:2021nsg, DarkSide-50:2022qzh,DarkSide:2022knj, SENSEI:2023zdf, PandaX:2024muv, DAMIC-M:2025luv} and the DM-SM interaction has to be governed by some {\it New Physics}~(NP)~\cite{Essig:2011nj}. However, in the lower mass scales, the NP must be weakly or selectively coupled to the SM to evade the experimental constraints. A theoretically well-motivated approach is to extend the SM gauge group $\mathcal{G}_{\rm SM}\equiv SU(3)_C\otimes SU(2)_L\otimes U(1)_Y$ with an additional $U(1)$ symmetry~\cite{Langacker:2008yv}. An SM-singlet state, non-trivially charged under the new abelian gauge group, can serve as a possible DM candidate with the associated neutral gauge boson $Z^\prime$ mediating the DM-SM interactions. Such abelian extensions of $\mathcal{G}_{\rm SM}$ can originate from the Grand Unified Theories~(GUT)~\cite{Robinett:1982tq,Langacker:1984dc}, extra-dimensional models~\cite{Antoniadis:1990ew,Appelquist:2000nn} or string compactifications~\cite{Goodsell:2010ie}. However, for the sub-GeV DM, a phenomenologically useful abelian extension can be formulated from the accidental global symmetries of the SM. The classical SM Lagrangian possesses a global symmetry ensuring the conservation of the baryon number $B$ and the individual lepton numbers $L_i$~[$i=e,\,\mu,\,\tau$]. In the perturbative regime, it can be featured by $\mathcal{G}_{LB}\supset U(1)_{B+L}\times U(1)_{B-L}\times U(1)_{L_\mu-L_\tau} \times U(1)_{L_\mu+L_\tau-2L_e}$~\cite{Heeck:2016xwg}~(here, $L=\sum_{i=e,\,\mu,\,\tau}L_i$), indicating a possibility to promote the difference between any two lepton numbers to a gauge quantum number~\cite{Foot:1990mn,He:1991qd,Foot:1994vd}. The corresponding $Z^\prime$ being naturally leptophilic, $\mathcal{G}_{\rm SM}\otimes U(1)_{L_i-L_j}$ provides a feasible BSM framework to study the DD prospects of a sub-GeV DM through the DM-electron scattering~\cite{Essig:2011nj,Graham:2012su,Essig:2015cda,Foldenauer:2018zrz,Dutta:2019fxn,Knapen:2021run,Hochberg:2021pkt,Kahn:2021ttr}. However, the continuous null results from the DD experiments are alarming and demand a respeculation. Models with an extended Higgs sector can lead to a destructive interference between the DM-nucleon scattering amplitudes corresponding to the light and heavy CP-even Higgs exchanges, resulting in a cancellation in the spin-independent scattering cross section. Such {\it blind spots} in the parameter space can't be detected by the DD searches~\cite{He:2011gc,Cheung:2012qy,Chang:2017gla,Huang:2014xua,Badziak:2015exr,Crivellin:2015bva,Han:2016qtc,Altmannshofer:2019wjb}. Similarly, Ref.~\cite{Gross:2017dan} has shown that in a simple extension of the SM with a complex scalar, a softly broken symmetry might ensure that the DD cross section vanishes at the tree-level for a Higgs-portal DM. A more generic approach has been proposed in Ref.~\cite{Das:2020ozo} considering a Higgs-portal WIMP model. Extending the SM quark sector with a 6-dimensional effective operator results in a negligible DM-nucleon scattering cross section over the entire parameter space. However, these cancellation mechanisms mostly work for the WIMPs scattering through the scalar or Higgs portals. Although a similar cancellation technique has been discussed in Ref.~\cite{Cai:2021evx} for a vector-portal DM within a generic abelian extension of the SM, the mechanism loses its charm once the kinetic mixing is incorporated. 

The present paper has considered a simple abelian extension of the SM gauge group with $U(1)_{L_\mu-L_\tau}$ where an SM-singlet vector-like lepton~(VLL) with a non-trivial $U(1)_{L_\mu-L_\tau}$ charge plays the role of the DM. An imposed $Z_2$ symmetry protects the DM once the $U(1)_{L_\mu-L_\tau}$ is spontaneously broken. This minimal setup for a leptophilic DM has been explored for a long time~\cite{Foldenauer:2018zrz,Asai:2020qlp, Hapitas:2021ilr, Figueroa:2024tmn,Wang:2025kit} and is particularly useful for studying the DM-electron scattering. However, it falls short to explain the DD bounds for the heavier DM masses\,$\sim \mathcal{O}(1-10^3)$ GeV. Hence, the paper has proposed a {\it semi-simple} augmentation of the minimal $\mathcal{G}_{\rm SM}\otimes U(1)_{L_\mu-L_\tau}$ particle content with additional VLLs and $SU(2)_L$-singlet neutral and charged scalars. Vector-like leptons are theoretically well-motivated BSM candidates to construct anomaly-free UV-complete theories and have been extensively studied in the literature~\cite{delAguila:2008pw, Joglekar:2012vc, Dermisek:2013gta, Carmona:2014iwa, Bhattacharya:2015qpa, Kumar:2015tna, Bhattacharya:2018fus, Bissmann:2020lge, DeJesus:2020yqx, Crivellin:2020ebi,Chakraborty:2021tdo, Guedes:2021oqx, De:2021crr, Cherchiglia:2021syq, Lee:2022nqz,Hamaguchi:2022byw,De:2023sqa,Cingiloglu:2024vdh,Dubey:2025sfh}. In the extended $\mathcal{G}_{\rm SM}\otimes U(1)_{L_\mu-L_\tau}$ framework, the DD cross section identically vanishes for both of the DM-electron and DM-quark scatterings, leading to a generic explanation for the continuous null results in the DD experiments. 

As an added advantage, the model can generate BSM corrections to certain leptonic observables, e.g., the lepton anomalous magnetic moments and $Z\ell^+\ell^-$~[$\ell=e,\,\mu,\,\tau$] couplings. The minimal $U(1)_{L_\mu-L_\tau}$ theory specifically affects only the muon and tau sectors at the one-loop level through the $Z^\prime$ exchange, whereas in the presence of the NP fields~(VLLs and BSM scalars), $(g-2)_e$ and $Z\to e^+e^-$ partial width can also be modified. In the SM, particularly, the $(g-2)_\mu$ has been predicted with a high computational precision, including the EW and hadronic contributions~\cite{Aliberti:2025beg}, while the ongoing experiments have reached a significant sensitivity to test it~\cite{Muong-2:2025xyk}. Thus, the muon anomalous magnetic moment acts as a critical observable to constrain the SM at the quantum level. Though the recent updates on the $(g-2)_\mu$ disfavor the need for a $\mu$-philic NP, beyond the current experimental sensitivity, the BSM contributions can still exist. Further, for the $(g-2)_e$, there is an observed discrepancy between its predicted~\cite{Aoyama:2019ryr} and measured values~\cite{Fan:2022eto} demanding an extension of the SM. However, note that the SM prediction for $(g-2)_e$ using the data driven method has an ambiguity that might be resolved with future experiments only. Although the SM predicts for the $(g-2)_\tau$~\cite{Eidelman:2007sb} as well, the present experiments are not sensitive enough to test/falsify the corresponding BSM contributions~\cite{DELPHI:2003nah}. Therefore, the flavor-specific NP can be a viable possibility in the lepton sector. The proposed extension of the minimal $U(1)_{L_\mu-L_\tau}$ model possesses different parameter spaces for the three different lepton flavors and thus, enhances the theoretical flexibility. However, the DM phenomenology, particularly the condition for a negligible DD cross section, plays a crucial role in correlating them. Moreover, the same parameters are responsible for inducing one-loop corrections to the $Z\to\ell^+\ell^-$ processes. Thus, the experimental bounds on the decay of the $Z$ boson to the charged SM leptons can be a vital tool to constrain the proposed NP interactions. Note that in the present paper, $Z^\prime$ exchange is the only way to generate one-loop corrections to the $\tau$-specific observables. It's worth emphasizing that using the bounds on the $Z\to\tau^+\tau^-$ decay, one can obtain a new exclusion limit on the minimal $U(1)_{L_\mu-L_\tau}$ parameter space. In the heavier $Z^\prime$ regime~$\left(\geq\mathcal{O}(100)~{\rm GeV}\right)$, a significant part of the parameter space, that was previously allowed by the experiments~\cite{CCFR:1991lpl}, can be excluded in the present work. 
 
The observation of neutrino oscillations~\cite{Super-Kamiokande:1998kpq} is another remarkable evidence of some NP beyond the SM. Experiments show a peculiar mixing pattern~(close to maximal between the $2^{\rm nd}$ and $3^{\rm rd}$ generations, and minimal between the $1^{\rm st}$ and $3^{\rm rd}$) with extremely light mass eigenstates~($\sim\mathcal{O}(10^{-2})$ eV) in the neutrino sector. Thus, due to its $\mu-\tau$ flavor symmetry, $U(1)_{L_\mu-L_\tau}$-extension is particularly useful to address the neutrino mixing parameters~\cite{Xing:2015fdg}. Therefore, the considered particle spectrum can be extended with three heavy right-handed neutrinos~(RHNs) to generate non-zero tiny neutrino masses and mixing among the lepton flavors through the Type-I seesaw mechanism~\cite{Minkowski:1977sc, Yanagida:1980xy,Mohapatra:1979ia,Schechter:1980gr}. The $U(1)_{L_\mu-L_\tau}$ symmetry breaking scale plays a significant role in explaining the neutrino oscillation data.

The rest of the paper has been organized as follows. Sec.~\ref{sec:2} presents a detailed discussion on the kinetic mixing in a generic $\mathcal{G}_{\rm SM}\otimes U(1)_X$ theory, followed by an explicit calculation of the loop-induced kinetic mixing terms. The extended $U(1)_{L_\mu-L_\tau}$ framework leading to a vanishing DM-electron and DM-quark scattering amplitude has been introduced in Sec.~\ref{sec:3} along with the associated experimental and theoretical constraints on the considered NP parameters. The phenomenology of a viable DM candidate has been explored in Sec.~\ref{sec:4}, while the Sec.~\ref{sec:5} has been dedicated to constraining the NP Yukawa couplings as well as the $U(1)_{L_\mu-L_\tau}$ gauge parameters through the $Z\to\ell^+\ell^-$ decays. The BSM contributions to the lepton anomalous magnetic moments have been elaborated in Sec.~\ref{sec:6}. Sec.~\ref{sec:neu} explains the origin of small neutrino masses and mixing angles, followed by the conclusion in Sec.~\ref{sec:7}.
\section{Kinetic Mixing}
\noindent
 \label{sec:2}
An abelian extension of the SM gauge group can be significant from the theoretical perspective to explain various BSM phenomena. Intriguingly, a notable feature of such theories is embodied in the vector portal coupling between the abelian gauge bosons of the $\mathcal{G}_{\rm SM}\otimes U(1)_X$ formulation. The interaction is naturally governed by the principles of Lorentz invariance and gauge symmetry, and can be expressed through the following renormalizable operator:
  \begin{align}
  \mathcal{L}_{\rm kin}= -\frac{\epsilon}{2}\,\,F^{\mu\nu}\mathbb{Z}_{\mu\nu}~.
  \label{eq:kin}
  \end{align}
Note that at this stage, $U(1)_X$ is a generic representation of the {\it new} abelian gauge group. The general results developed in this section will later be used for studying the phenomenology of $U(1)_{L_\mu-L_\tau}$. 
 \begin{figure}[!ht]
 \centering
 \subfloat[(a)]{\includegraphics[scale=0.7]{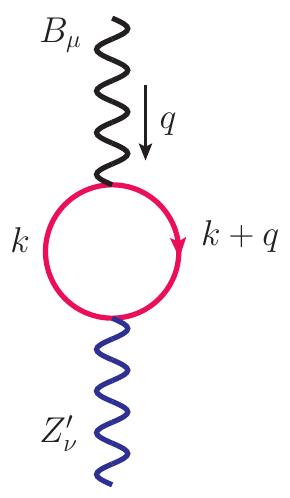}}\qquad\qquad\qquad
 \subfloat[(b)]{\includegraphics[scale=0.7]{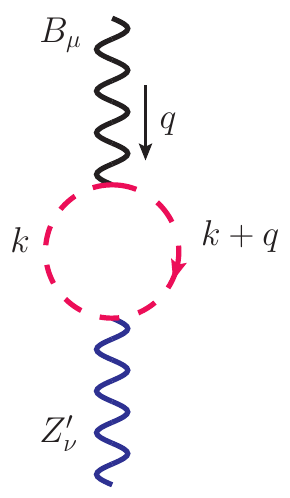}}\qquad\qquad
 \subfloat[(c)]{\includegraphics[scale=0.7]{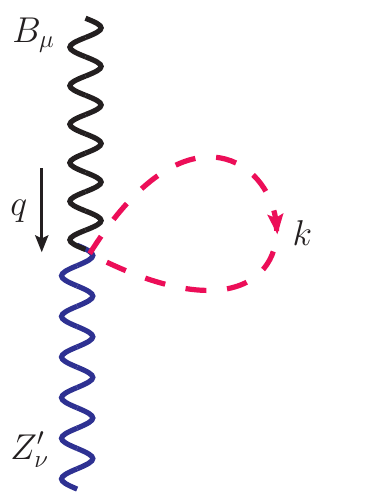}}
 \caption{Kinetic mixing at one-loop level due to a generic fermion $f$~(a) and a scalar $S$~(b, c) with non-zero $X$, $Y$ charges. $B_\mu$ and $Z^\prime_\nu$ are the gauge bosons associated with $U(1)_Y$ and $U(1)_X$, respectively. $q$ defines the transferred momentum.}
 \label{fig:kin1}
 \end{figure}
In Eq.~\ref{eq:kin}, $F^{\mu\nu}$ and $\mathbb{Z}^{\mu\nu}$ represent the field strength tensors corresponding to the hypercharge and the $U(1)_X$ gauge bosons, respectively. The dimensionless coupling constant $\epsilon$ is commonly known as the kinetic mixing, and in general, can be a fundamental parameter in the theory. Though it is not protected by any symmetry, its value crucially depends on the UV-completion of $U(1)_X$~\cite{He:1991qd, Bauer:2018onh}. If the new abelian theory can be embedded into a higher dimensional non-abelian gauge group~($G_X$) which doesn't possess any mixing with the SM during the entire symmetry breaking chain~$\big($i.e., $G_X\to U(1)_X\big)$, it is guranteed that the kinetic mixing between the $U(1)_X$ and $U(1)_Y$ gauge bosons will be finite. $U(1)$ theories obtained by gauging the difference of lepton numbers fall into this category. It is straightforward to embed $U(1)_{L_i-L_j}$~($i,\, j=e,\, \mu,\, \tau$) into $SU(2)_{L_i-L_j}$, and break it to the gauge boson associated with its diagonal generator $\sigma_3={\rm diag}\,(1,\,-1)$~\cite{He:1991qd,Heeck:2011wj}. However, the argument doesn't work for $U(1)_{B-L}$, where the gauge couplings of the possible non-abelian parent group can't be independent of the gauge couplings of $U(1)_Y$, resulting in a renormalization scale-dependent kinetic mixing. 

Assuming the $U(1)_X$ hails from the former class, one can set $\epsilon_{\rm \,tree}=0$. However, in the presence of the fields that are simultaneously charged under $U(1)_X$ and $U(1)_Y$, there exists an unavoidable loop-induced contribution to Eq.~\eqref{eq:kin}. With $\epsilon_{\rm \,tree}=0$, the leading order kinetic mixing arises at the one-loop level and can be phenomenologically vital for the present work. Therefore, the rest of this section has been dedicated to building up the analytical expressions of the loop-induced kinetic mixings for a generic fermion $f$ and a scalar $S$, transforming non-trivially under both of the abelian gauge groups. Note that it is just for the sake of completeness. One can easily find the calculations in Ref.~\cite{Peskin:1995ev}.
 \subsection{Fermion Loop}
 Fig.~\ref{fig:kin1}\,(a) shows the generation of kinetic mixing at the one-loop order in the presence of a fermionic field $f$, carrying non-zero abelian charges under the hypercharge group and the $U(1)_X$. If $g_1$ and $g^\prime$ represent the gauge couplings corresponding to $U(1)_Y$ and $U(1)_X$, respectively, the covariant derivative for $f$ can be defined as,
 \begin{align}
 D_\mu=\partial_\mu-ig_1Y_fB_\mu-ig^\prime Q_fZ^\prime_\mu~.
 \label{eq:cov}
 \end{align}
 Here, $Y_f$ and $Q_f$ are the abelian charges of $f$, and $Z^\prime$ denotes the neutral gauge boson associated with the $U(1)_X$ theory. The loop contribution from Fig.~\ref{fig:kin1}\,(a) can be formulated as,
 \begin{align}
 i\,\Pi^{\mu\nu}(q)&=(-1)\int\frac{d^4k}{(2\pi)^4}\,\,{\rm Tr}\Bigg[(-ig_1Y_f\gamma^\mu)\,\frac{i(\slashed{k}+\slashed{q}+m_f)}{(k+q)^2-m_f^2}\, (-ig^\prime Q_f\gamma^\nu)\,\frac{i(\slashed{k}+m_f)}{k^2-m_f^2}\Bigg]\nonumber\\
 &=-4g_1g^\prime Y_fQ_f\int\frac{d^4k}{(2\pi)^4}\Bigg[\frac{k^\mu(k+q)^\nu+k^\nu(k+q)^\mu-g^{\mu\nu}\{k.(k+q)-m_f^2\}}{(k^2-m_f^2)\{(k+q)^2-m_f^2\}}\Bigg]
 \label{eq:floop1}
 \end{align}
 Using Feynman parametrization, Eq.~\eqref{eq:floop1} can be recast as,
 \begin{align}
 i\,\Pi^{\mu\nu}(q)&=-4g_1g^\prime Y_fQ_f\int_0^1 dx\int\frac{d^4P}{(2\pi)^4}\Bigg[\frac{1}{(P^2-\Delta_f)^2}\Bigg]\nonumber\\
 &\qquad\qquad\times\Big[2P^\mu P^\nu-g^{\mu\nu}P^2-2x(1-x)q^\mu q^\nu+g^{\mu\nu}\{m_f^2+x(1-x)q^2\}\Big]~,
 \label{eq:floop2}
 \end{align}
 where, $\Delta_f=m_f^2-x(1-x)q^2$. The momentum integration reduces Eq.~\eqref{eq:floop2} to,
 \begin{align}
 i\,\Pi^{\mu\nu}(q)&=i(q^2g^{\mu\nu}-q^\mu q^\nu)\,\epsilon_f(q^2)~.
 \end{align}
 Thus, the kinetic mixing corresponding to Fig.~\ref{fig:kin1}\,(a) can be defined as,
 \begin{align}
 \epsilon_f(q^2)=\frac{g_1g^\prime Y_fQ_f}{2\pi^2}\int_0^1 dx~ x(1-x)\,\ln\Bigg[\frac{m_f^2-x(1-x)q^2}{\Lambda^2}\Bigg]~.
 \end{align}
 Here, $\Lambda$ represents the renormalization scale. Further, for a vanishingly small momentum transfer, one obtains,
 \begin{align}
 \epsilon_f(0)=\frac{g_1g^\prime Y_fQ_f}{12\pi^2}\,\ln\left(\frac{m_f^2}{\Lambda^2}\right)~.
 \label{eq:floop3}
 \end{align}
 
 \subsection{Scalar Loop}
 In the presence of a scalar $S$, charged under both the abelian gauge groups, the total one-loop contribution to the kinetic mixing can be computed as a sum of the individual contributions arising from Figs.~\ref{fig:kin1}\,(b) and \ref{fig:kin1}\,(c). Note that Eq.~\eqref{eq:cov} is equally applicable for $S$, with appropriate abelian charges, i.e., $Y_S$ and $Q_S$. Thus,
 \begin{align}
 i\,\Pi^{\mu\nu}(q)=i\,\Pi_{(a)}^{\mu\nu}(q)+i\,\Pi_{(b)}^{\mu\nu}(q)~,
 \end{align}
 where,
 \begin{align}
 i\,\Pi_{(a)}^{\mu\nu}(q)&=\int\frac{d^4k}{(2\pi)^4}\,\Bigg[(-ig_1Y_S)\,\frac{i(2k+q)^\mu}{(k+q)^2-M_S^2}\, (-ig^\prime Q_S)\,\frac{i(2k+q)^\nu}{k^2-M_S^2}\Bigg]\nonumber\\
 &=g_1g^\prime Y_SQ_S\int\frac{d^4k}{(2\pi)^4}\,\Bigg[\frac{(2k+q)^\mu (2k+q)^\nu}{(k^2-M_S^2)\{(k+q)^2-M_S^2\}}\Bigg]~,\\
 i\,\Pi_{(b)}^{\mu\nu}(q)&=-2g_1g^\prime Y_SQ_S\,\,g^{\mu\nu}\int\frac{d^4k}{(2\pi)^4}\,\Bigg[\frac{1}{k^2-M_S^2}\Bigg]~.
 \end{align}
 Therefore,
 \begin{align}
 i\,\Pi^{\mu\nu}(q)=&~g_1g^\prime Y_SQ_S\int\frac{d^4k}{(2\pi)^4}\,\Bigg[\frac{1}{(k^2-M_S^2)\{(k+q)^2-M_S^2\}}\Bigg]\nonumber\\
 &\qquad\qquad\qquad\times \Big[4k^\mu k^\nu+2k^\mu q^\nu +2k^\nu q^\mu +q^\mu q^\nu-2g^{\mu\nu}\{(k+q)^2-M_S^2\}\Big]~.
 \label{eq:sloop1}
 \end{align}
 With Feynman parametrization, Eq.~\eqref{eq:sloop1} can be reformulated as,
 \begin{align}
 i\,\Pi^{\mu\nu}(q)=&~g_1g^\prime Y_SQ_S\int_0^1dx\int\frac{d^4P}{(2\pi)^4}\,\Bigg[\frac{1}{(P^2-\Delta_S)^2}\Bigg]\nonumber\\
 &\qquad\qquad\times \Big[4P^\mu P^\nu-2g^{\mu\nu}P^2 -2g^{\mu\nu}\{(1-x)^2q^2-M_S^2\}+(1-2x)^2q^\mu q^\nu\Big]\nonumber\\
 =&~i(q^2g^{\mu\nu}-q^\mu q^\nu)\,\epsilon_S(q^2)~.
 \label{eq:sloop2}
 \end{align}
 Here $\Delta_S=M_S^2-x(1-x)q^2$. Note that Eq.~\eqref{eq:sloop2} has been obtained through the momentum integral followed by a simplification due to the symmetry of the Feynman integral about $x=1/2$. One can easily check that,
 \begin{align}
 \lim_{d\to\, 4}\int_0^1 dx \Bigg[\frac{1-2x}{(\Delta_S)^{2-d/2}}\Bigg]=0~.
\end{align}  
 The renormalized scalar contribution to the kinetic mixing is given by,
 \begin{align}
 \epsilon_S(q^2)=-\frac{g_1g^\prime Y_SQ_S}{8\pi^2}\int_0^1dx~ x(1-2x)\,\ln\Bigg[\frac{M_S^2-x(1-x)q^2}{\Lambda^2}\Bigg]~.
 \label{eq:sloop3}
\end{align}  
As before, $\Lambda$ stands for the renormalization scale. For the limiting case of $q^2\to 0$, Eq.~\eqref{eq:sloop3} becomes,
\begin{align}
 \epsilon_S(0)=\frac{g_1g^\prime Y_SQ_S}{48\pi^2}\,\ln\left(\frac{M_S^2}{\Lambda^2}\right)~.
 \label{eq:sloop4}
\end{align} 

Note that if $f$ and/or $S$ appear as multiplets under some exact non-abelian gauge symmetry, one must incorporate a degeneracy factor in Eq.~\eqref{eq:floop3} and/or \eqref{eq:sloop4}, respectively. For example, if $f$ is a color-triplet, an extra  multiplicative factor of $3$ must be considered in Eq.~\eqref{eq:floop3}. 
\section{The Model}
 \noindent
 \label{sec:3}
 As indicated in the Introduction, the prime goal of the present paper is to correlate the null results in the direct detection experiments to the NP contributing to the lepton sector. The ongoing DD searches are fundamentally based on the interaction of the DM with either quarks or electrons, with the detector sensitivity being strongly dependent on the chosen mass regime of the DM candidate. However, till now, no positive signal has been reported from the experiments. The $U(1)_{L_\mu-L_\tau}$-extension of the SM is a naturally preferable framework to accommodate the DD results, as the viable DM candidate~(being an SM-singlet) can have only loop-suppressed interactions with the quarks and the electron. At the leading order, the DM-nucleus or DM-electron scattering cross-section will be proportional to $|\epsilon(0)|^2$. In the minimal $\mathcal{G}_{\rm SM}\otimes U(1)_{L_\mu-L_\tau}$, 
 \begin{align}
 \epsilon(0)=\frac{g_1g^\prime}{12\pi^2}\,\ln\left(\frac{m_\tau^2}{m_\mu^2}\right)=\frac{e g^\prime}{6\pi^2 \cos\theta_W}\,\ln\left(\frac{m_\tau}{m_\mu}\right)~,
 \label{eq:kin_sm}
 \end{align}
 with $e$ and $\theta_W$ being the electronic charge and weak mixing angle, respectively. One can trivially evaluate Eq.~\eqref{eq:kin_sm} to obtain $|\epsilon(0)/g^\prime|^2=2.7\times 10^{-4}$. Further, the non-relativistic~(NR) DM-target scattering cross section for a $Z^\prime$-portal DM can be formulated as,
 \begin{align}
 \sigma_{{\rm DM}-T}= \frac{e^2\mathbb{M}^2_{{\rm DM}-T}}{\pi \cos^2\theta_W}\times\left(\frac{g^\prime}{M_{Z^\prime}}\right)^4\times\left|\frac{\epsilon(0)}{g^\prime}\right|^2~,
 \label{eq:sig_DD}
 \end{align}
with $T$ being either nucleon~($N$) or electron~($e$). $\mathbb{M}_{{\rm DM}-T}$ is the reduced mass of the DM-target 2-body system. Let's fix $(g^\prime/M_{Z^\prime})\sim \mathcal{O}(10^{-2})$ GeV$^{-1}$ for the numerical estimation of $\sigma_{{\rm DM}-T}$. Thus, with the existing DD bounds from the DM-nucleus scattering experiments~\cite{LZ:2022lsv}, Eq.~\eqref{eq:sig_DD} leads to $|\epsilon/g^\prime|^2<10^{-9}$ for a DM mass of 1 TeV. However, for a sub-GeV dark matter, a much relaxed upper bound can be obtained corresponding to the DD constraints~\cite{PandaX:2024muv, DAMIC-M:2025luv}\,\footnote{For a recent analysis, one can go through Ref.~\cite{Cheek:2025nul}.} and the white dwarf observations~\cite{Dasgupta:2019juq, Bell:2021fye}. Note that in the sub-GeV mass regime, the most stringent bounds on $|\epsilon/g^\prime|$ come from the experiments searching for a hidden gauge sector and will be discussed in Sec.~\ref{subsec:2C}. Therefore, even though the minimal setup can explain the present DD constraints for DM-electron scattering, it completely fails in the WIMP mass regime. 
 
The paper has proposed an extension of the minimal $\mathcal{G}_{\rm SM}\otimes U(1)_{L_\mu-L_\tau}$ particle spectrum with three $SU(2)_L$-singlet vector-like leptons~(VLL) $\psi$ and $\chi_{1,\,2}$, a complex $SU(2)_L$-singlet scalar $\eta$, and a real SM-singlet scalar $\xi$. Vector-like fermions transform in the non-chiral representations of the unbroken $\mathcal{G}_{\rm SM}$ and are particularly suitable to maintain the anomaly-free structure of the considered gauge theory. Further, an extra $Z_2$ symmetry has been imposed to regulate the tree-level interactions of these five NP fields. $\chi_1$ has been assumed to be the lightest SM-singlet $Z_2$-odd state such that it can be a viable DM candidate in the present BSM formulation. Moreover, three right-handed Majorana neutrinos~($N_R^\ell$), all even under the imposed $Z_2$ symmetry, have been considered to explain the neutrino mass and mixing angles using the Type-I seesaw mechanism. For convenience, the proposed framework can be called as {\it Beyond the Minimal $U(1)_{L_\mu-L_\tau}$} model or in short, BM-$U(1)_{L_\mu-L_\tau}$. Table~\ref{tab:parti} enlists the complete particle content of the model with their gauge quantum numbers. Note that the main role of the SM-singlet scalar $\phi$ is to break the $U(1)_{L_\mu-L_\tau}$ symmetry spontaneously with a vacuum expectation value~(VEV) $v^\prime$ and it is a part of the minimal $\mathcal{G}_{\rm SM}\otimes U(1)_{L_\mu-L_\tau}$ model. However, with a proper charge assignment, it can generate mixing among the different neutrino flavors.
  \begin{table}[!ht]
\begin{tabular}{|c|c|c|c|}
\hline
Fields & Spin & $SU(3)_C\otimes SU(2)_L\otimes U(1)_Y\otimes U(1)_{L_\mu-L_\tau}$ & $Z_2$\\
\hline
\hline
$L_L^e=(\nu_e\quad e)^T$  & $1/2$ & ({\bf 1}, {\bf 2}, $-1/2$, $0$) & $+1$ \\
$L_L^\mu=(\nu_\mu\quad \mu)^T$  & $1/2$ & ({\bf 1}, {\bf 2}, $-1/2$, $1$) & $+1$ \\
$L_L^\tau=(\nu_\tau\quad \tau)^T$ & $1/2$ & ({\bf 1}, {\bf 2}, $-1/2$, $-1$) & $+1$ \\
		$ e_R$ & $1/2$ & ({\bf 1}, {\bf 1}, $-1$, $0$) & $+1$\\
		$\mu_R$ & $1/2$ & ({\bf 1}, {\bf 1}, $-1$, $1$) & $+1$\\
		$\tau_R$ & $1/2$ & ({\bf 1}, {\bf 1}, $-1$, $-1$) & $+1$\\
		$Q_L=(u_L\quad d_L)^T$ & $1/2$ & ({\bf 3}, {\bf 2}, $1/6$, 0) & $+1$\\
		$U_R=(u_R,\,c_R,\,t_R)$ & $1/2$ & ({\bf 3}, {\bf 1}, $2/3$, 0) & $+1$\\
		$D_R=(d_R,\,s_R,\,b_R)$ & $1/2$ & ({\bf 3}, {\bf 1}, $-1/3$, 0) & $+1$\\
		$H = (H^+ \quad H^0)^T $ & $0$ & ({\bf 1}, {\bf 2}, $1/2$, 0)  & $+1$\\
		$\phi$ & $0$ & ({\bf 1}, {\bf 1}, 0, $Q_\phi$)	& $+1$\\
\hline
		$\eta$ & $0$ & ({\bf 1}, {\bf 1}, $Y_\eta$, $Q_\eta$) & $-1$\\
		$\xi$ & $0$ & ({\bf 1}, {\bf 1}, $0$, $0$) & $-1$\\
		$(\chi_1)_{L,\,R}$ & $1/2$ & ({\bf 1}, {\bf 1}, 0, $Q_1$)	& $-1$\\
		$(\chi_2)_{L,\,R}$ & $1/2$ & ({\bf 1}, {\bf 1}, 0, $Q_2$)	& $-1$\\
		$\psi_{L,\,R}$  & $1/2$ & ({\bf 1}, {\bf 1}, $Y_\psi$, $Q_\psi$) & $-1$ \\
		$N_R^e$ & $1/2$ & ({\bf 1}, {\bf 1}, $0$, $0$) & $+1$ \\
		$N_R^\mu$ & $1/2$ & ({\bf 1}, {\bf 1}, $0$, $1$) & $+1$ \\
		$N_R^\tau$ & $1/2$ & ({\bf 1}, {\bf 1}, $0$, $-1$) & $+1$ \\
		\hline
\end{tabular} 
\caption{The fields and their transformations under $\mathcal{G}_{\rm SM}\otimes U(1)_{L_\mu-L_\tau}\otimes Z_2$. After electroweak symmetry breaking~(EWSB), the electromagnetic~(EM) charge can be defined as $\mathbb{Q}_{\rm EM}=T_3+Y$. The upper block represents the particle content for a minimal $\mathcal{G}_{\rm SM}\otimes U(1)_{L_\mu-L_\tau}$ theory, whereas the lower block contains the additional fields required to construct the BM-$U(1)_{L_\mu-L_\tau}$.}
\label{tab:parti}
\end{table}

Apart from the $2^{\rm nd}$ and $3^{\rm rd}$ generation SM leptons, $\psi$ and $\eta$ are the only fields charged under $U(1)_Y$ and $U(1)_{L_\mu-L_\tau}$, simultaneously. Thus, with the augmented particle spectrum, the total one-loop contribution to the kinetic mixing parameter at the $q^2\to 0$ limit can be defined as,
\begin{align}
\epsilon_{\mu\tau}(0)=&~\frac{g_1g^\prime}{12\pi^2}\,\ln\left(\frac{m_\tau^2}{m_\mu^2}\right)+\frac{g_1g^\prime Y_\psi Q_\psi}{12\pi^2}\,\ln\left(\frac{m_\psi^2}{\Lambda_1^2}\right)+\frac{g_1g^\prime Y_\eta Q_\eta}{48\pi^2}\,\ln\left(\frac{M_\eta^2}{\Lambda_1^2}\right)\nonumber\\
=&~\frac{g_1g^\prime}{12\pi^2}\,\ln\left(\frac{m_\tau^2}{m_\mu^2}\right)+\frac{g_1g^\prime}{12\pi^2}\Bigg[Y_\psi Q_\psi\,\ln\left(\frac{m_\psi^2}{\Lambda_1^2}\right)+\frac{Y_\eta Q_\eta}{4}\,\ln\left(\frac{M_\eta^2}{\Lambda_1^2}\right)\Bigg]~,
\label{eq:kin_fin}
\end{align}
where $m_\psi$ and $M_\eta$ denote the masses of $\psi$ and $\eta$, respectively. $\Lambda_1$ specifies the renormalization scale associated with the BSM fields. The finiteness of $\epsilon_{\mu\tau}(0)$ demands 
\begin{align}
Y_\psi Q_\psi=-\frac{Y_\eta Q_\eta}{4}~,
\label{eq:fin}
\end{align}
resulting in,
\begin{align}
\epsilon_{\mu\tau}(0)=&~\frac{g_1g^\prime}{12\pi^2}\Bigg[\ln\left(\frac{m_\tau^2}{m_\mu^2}\right)+Y_\psi Q_\psi\,\ln\left(\frac{m_\psi^2}{M_\eta^2}\right)\Bigg]\nonumber\\
=&~\frac{g_1g^\prime}{12\pi^2}\,\ln\Bigg[\left(\frac{m_\tau^2}{m_\mu^2}\right)\times\left(\frac{m_\psi^2}{M_\eta^2}\right)^{Y_\psi Q_\psi}\Bigg]~.
\end{align}
Clearly, with these new degrees of freedom, an exact cancellation condition for the leading order kinetic mixing term can be obtained as,
\begin{align}
\frac{m_\psi}{M_\eta}=\left(\frac{m_\mu}{m_\tau}\right)^{1/Y_\psi Q_\psi}~.
\label{eq:cancel1}
\end{align}
However, $Y_\psi$ and $Q_\psi$~(or equivalently $Y_\eta$ and $Q_\eta$) being arbitrary parameters, Eq.~\eqref{eq:cancel1} doesn't represent a unique condition and one must incorporate other NP interactions to determine the gauge charges unambiguously. 

With the enlarged particle spectrum, let's consider the following Yukawa interactions:
\begin{align}
\mathcal{L}\supset -\Big[y_e\,\bar{e}_R\eta\chi_2+y_\mu\,\bar{\mu}_R\xi\psi+{\rm h.c.}\Big]~.
\label{eq:emu1}
\end{align}
Thus, Eq.~\eqref{eq:emu1} has coupled $\eta$ and $\psi$ with two different lepton flavors. The fields having non-zero hypercharges result in one-loop corrections to the $Z\to e^+e^-$ and $Z\to \mu^+\mu^-$ decays, which can be crucial to constrain the NP Yukawa interactions. Moreover, the first term can generate a BSM contribution to $(g-2)_e$, whereas the second term results in an additional NP contribution to $(g-2)_\mu$ within the BM-$U(1)_{L_\mu-L_\tau}$ model.  

The charge neutrality of the Yukawa terms demands,
\begin{align}
&Y_\eta=-1\,,\qquad\qquad Q_2=-Q_\eta~,\nonumber\\
&Y_\psi=-1\,,\qquad\qquad Q_\psi=1~.
\label{eq:YQ}
\end{align} 
Further, combining Eqs.~\eqref{eq:fin} and \eqref{eq:YQ}, one obtains
\begin{align}
Q_\eta=-4\quad\Rightarrow \quad Q_2=4~.
\end{align}
Thus, with the present charge assignments, $\eta$ and $\psi$ in Table~\ref{tab:parti} should be read as $\eta^-$ and $\psi^-$, defining a singly charged scalar and VLL, respectively. Moreover, Eq.~\eqref{eq:cancel1} can now be recast as,
\begin{align}
\frac{m_\psi}{M_\eta}=\frac{m_\tau}{m_\mu}=16.8~.
\label{cancel2}
\end{align}
For all the subsequent discussions, this mass ratio will be followed between $\eta$ and $\psi$. At this stage, the singlet scalar $\phi$ can have any arbitrary non-zero $U(1)_{L_\mu-L_\tau}$ charge $Q_\phi$, provided $Q_\phi\neq \pm|Q_1-Q_2|$. $Q_1$ is also a free parameter in the theory and can't be uniquely fixed through the considered phenomena. However, to forbid additional Yukawa terms, e.g., $\bar{\ell}_R\eta\chi_1$ and $\bar{\ell}_R\xi\chi_1$~[$\ell=e,\,\mu,\,\tau$], one can set $Q_1\neq 0,\,\pm 1,\, Q_2,\,-(Q_\eta\pm 1)$ for the present analysis. In the presence of such Yukawa interactions, $\chi_1$ could possess a non-negligible DD cross section with the leading order contribution arising from the photon and $Z$ penguin diagrams~\cite{Herrero-Garcia:2018koq}. Note that Eq.~\eqref{eq:emu1} is not a unique NP construction to fix the gauge charges of $\eta$ and $\psi$. As an alternative choice, one could have considered a single Yukawa term involving $\eta$ and $\psi$, e.g., $\bar{e}_R\eta\psi$ or $\bar{\mu}_R\eta\psi$. However, such terms would lead to fractional EM charges for $\eta$ and $\psi$, resulting in stringent collider constraints~\cite{CMS:2024eyx}. Further, in such cases, an unnatural mass hierarchy can occur between $\eta$ and $\psi$, making $\psi$ too heavy to be probed through the future colliders. Thus, Eq.~\eqref{eq:emu1} represents the minimal experimentally feasible extension of $\mathcal{G}_{\rm SM}\otimes U(1)_{L_\mu-L_\tau}$, required to explain the non-observation of the DM through the DD experiments with leptophilic NP interactions.   

Therefore, the complete Lagrangian for the proposed framework can be cast as,
\begin{align}
\mathcal{L}=\mathcal{L}_{\rm SM}+\mathcal{L}_{Z_2}+\mathcal{L}_{\rm N}-\mathbb{V}(H,\,\eta,\,\xi,\,\phi)~,
\end{align}
where $\mathcal{L}_{\rm SM}$ represents the SM sector only, while the NP interactions involving $Z^\prime$ and the $Z_2$-odd fields are encapsulated in 
\begin{align}
\mathcal{L}_{Z_2}=&~-\frac{1}{4}\,\mathbb{Z}^{\mu\nu}\mathbb{Z}_{\mu\nu}+(D^\mu\eta)^\dagger(D_\mu\eta)+\frac{1}{2}(\partial^\mu\xi)(\partial_\mu\xi)+(D^\mu\phi)^\dagger(D_\mu\phi)+\bar{\psi}(i\slashed{D}-m_\psi)\psi\nonumber\\
&+\sum_{j\,=\,1,\,2}\bar{\chi}_j(i\slashed{D}-m_j)\chi_j-\Big[y_e\,\bar{e}_R\eta\chi_2+y_\mu\,\bar{\mu}_R\xi\psi+{\rm h.c.}\Big]~.
\label{eq:NP}
\end{align}
$\mathcal{L}_{\rm N}$ defines the neutrino sector for the considered extension and will be discussed in Sec.~\ref{sec:neu}. The function $\mathbb{V}$ represents the scalar potential. 
Note that due to the proposed charge assignment of the BSM fields, the non-observation of the charged lepton flavor violating~(CLFV) processes~\cite{TheMEG:2016wtm,Aubert:2009ag, Bellgardt:1987du,Hayasaka:2010np} can be trivially explained. Further, to accommodate the experimental constraints on the electric dipole moment~(EDM) of leptons~\cite{ACME:2018yjb, Muong-2:2008ebm, Belle:2002nla}, one can assume $y_e$ and $y_\mu$ to be real without compromising any vital phenomenological input to the $(g-2)_\ell$, $Z\to\ell^+\ell^-$~[$\ell=e,\,\mu$] and the DM observables. In the Eq.~\eqref{eq:NP}, $m_{1,\,2}$ represent the mass terms for $\chi_1$ and $\chi_2$, respectively, with $\chi_1$ being the proposed DM candidate. The covariant derivative $D$ is defined as,
\begin{align}
D_\mu~&=\partial_\mu-ig_1 Y B_\mu-ig^\prime Q^\prime Z^\prime_\mu~,
\label{eq:codev}
\end{align}
where, $Q^\prime$ and $Y$ are generic notations for the $U(1)_{L_\mu-L_\tau}$ and $U(1)_Y$ charges of a particular field, respectively.  

The theoretical upper bound on the NP Yukawa couplings $y_e$ and $y_\mu$ comes from the perturbative unitarity~(PU). The tree-level PU bounds can be obtained from the $2\to 2$ scattering amplitude of the associated fields and are, in general, specific to the considered model. The angular dependence of the scattering amplitude can be eliminated by projecting it onto the partial waves of total angular momentum $J$. $J=0\,(1)$ corresponds to the scattering between two fermion fields with the same\,(opposite) helicity, whereas, in the present formulation, $J=1/2$ can only be associated with the scattering between a fermion and a scalar field~\cite{Urquia-Calderon:2024rzc}. For the considered Yukawa interactions, all the fermions and scalars are $SU(2)_L$-singlets. Moreover, $\xi$ is a real scalar, while $\eta^-$ is complex. Thus, following Ref.~\cite{Allwicher:2021rtd}, the PU bounds on $y_e$ and $y_\mu$ corresponding to different $J$ values have been listed in Table~\ref{tab:PU}. 
\begin{table}[!ht]
\begin{tabular}{|c|c|c|c|}
\hline
$\qquad {\rm Yukawa~ Coupling}\qquad$ & $\qquad J=0\qquad$ & $\qquad J=1/2\qquad$ & $\qquad J=1\qquad$\\
\hline\hline
$|y_e|$ & $<4\sqrt{\pi}$ & $<\sqrt{8\pi}$ & $<\sqrt{8\pi}$\\
\hline
$|y_\mu|$ & $<\sqrt{4\pi}$ & $<4\sqrt{\pi/3}$ & $<4\sqrt{\pi}$\\
\hline
\end{tabular}
\caption{Upper bounds on $y_e$ and $y_\mu$ corresponding to $J=0,\,1/2$ and $1$.}
\label{tab:PU}
\end{table}

Therefore, altogether, the most stringent theoretical constraints on $y_e$ and $y_\mu$ can be read as,
\begin{align}
|y_e|<\sqrt{8\pi}~\,,\qquad\qquad |y_\mu|<\sqrt{4\pi}~.
\label{eq:PU}
\end{align}
Evidently, the complex scalar singlet is theoretically more relaxed than its real counterpart.
\subsection{Scalar Sector}
The SM scalar sector being extended with a singly charged and two SM-singlet scalars, the potential term can be formulated as,
\begin{align}
\mathbb{V}(H,\,\eta,\,\xi,\,\phi)&=\mu_H^2(H^\dagger H)+\lambda_H(H^\dagger H)^2+\tilde{M}_{\eta}^2(\eta^* \eta)+\lambda_\eta(\eta^* \eta)^2+\frac{1}{2}\,\tilde{M}_{\xi}^2\,\xi^2+\lambda_\xi\,\xi^4\nonumber\\
&+\mu_\phi^2(\phi^*\phi)+\lambda_\phi(\phi^* \phi)^2+\lambda_1(H^\dagger H)(\eta^* \eta)+\lambda_2(H^\dagger H)\xi^2+\lambda_3(H^\dagger H)(\phi^* \phi)\nonumber\\
&+\lambda_4(\eta^* \eta)(\phi^* \phi)+\lambda_5(\phi^* \phi)\xi^2+\lambda_6(\eta^* \eta)\xi^2~.
\label{eq:V_pot}
\end{align}
It should be noted that the abelian gauge symmetries of the theory, along with the discrete $Z_2$, forbid any other $SU(2)_L$-invariant possibility in Eq.~\eqref{eq:V_pot}. In the scalar potential, $\tilde{M}_{\eta}$ and $\tilde{M}_{\xi}$ define the bare mass terms for $\eta$ and $\xi$, respectively. Further, one has to consider $\mu_H^2<0$ and $\mu_\phi^2<0$ so that $H$ and $\phi$ can acquire non-zero VEVs leading to spontaneous symmetry breaking~(SSB). Indeed, $\mathbb{V}$ represents a non-minimal renormalizable extension of the SM scalar sector and may improve the stability of the vacuum. However, the scalar potential has to be bounded from below to ensure the vacuum stability, i.e., $\mathbb{V}$ should not hit the negative infinity in any of the field directions. In the SM, it can be trivially achieved by considering $\lambda_H>0$. However, with the enhanced scalar degrees of freedom, one must adopt a rigorous mathematical treatment to derive the vacuum stability conditions. Note that the terms with dimensionful couplings don't play a role in the calculation of vacuum stability, as they contribute negligibly in the limit of large field values in comparison to the quartic terms of the scalar potential. Thus, the vacuum stability conditions can be directly obtained from the copositivity of the quartic coupling matrix~\cite{Kannike:2012pe}. In particular, for Eq.~\eqref{eq:V_pot}, if one considers a non-negative vector $\mathbf{X}=(\phi^* \phi\quad H^\dagger H\quad \eta^* \eta\quad \xi^2)^T$, the quartic part can be expressed as $\mathbf{X}^T\,\Omega\, \mathbf{X}$ with 
\begin{align}
\Omega~=~\left[\begin{array}{c c c c}
\lambda_\phi\qquad & \lambda_3/2\qquad & \lambda_4/2\qquad & \lambda_5/2\\
\lambda_3/2\qquad & \lambda_H\qquad & \lambda_1/2\qquad & \lambda_{2}/2\\
\lambda_4/2\qquad & \lambda_1/2\qquad & \lambda_\eta\qquad & \lambda_6/2\\
\lambda_5/2\qquad & \lambda_{2}/2\qquad & \lambda_6/2\qquad & \lambda_\xi
\end{array}\right]
\end{align}
defining the quartic coupling matrix. Ref.~\cite{PING1993109} provides the copositivity conditions for a generic $4\times 4$ symmetric matrix. Depending on the signs of the off-diagonal elements of $\Omega$, there can be eight different cases~\cite{SalimAdam:2025wlp}. However, it is easier to follow a more general criterion from the Cottle-Habetler-Lemke~(CHL) theorem~\cite{cottle1970classes} to test the copositivity of such large matrices. The theorem states that,\\
{\it If the order $n-1$ principal submatrices of a real symmetric matrix $A$ of order $n$ are copositive, then $A$ is copositive if and only if det$(A)\geq 0$ or some element(s) of Adj$(A)$ are negative.}\\
Here, the Adj$(A)$ signifies the adjugate of the matrix $A$. However, to have a strong vacuum  stability condition, one must follow the strict copositivity criteria, i.e., only the exact inequalities should be considered. For any real symmetric matrix of order $n$, the principal submatrices of order $n-1$ can be obtained by deleting its $k^{\rm th}$ row and $k^{\rm th}$ column simultaneously, where $k=1,\,\cdots,\,n$. Thus, there can be $n$ principal submatrices of order $n-1$ for a symmetric matrix of order $n$. Therefore, corresponding to $\Omega$, one can have $4$ principal submatrices of order $3$, as given below.
\begin{align}
\mathcal{M}_1=&~\left[\begin{array}{c c c }
\lambda_H\qquad & \lambda_1/2\qquad & \lambda_{2}/2\\
\lambda_1/2\qquad & \lambda_\eta\qquad & \lambda_6/2\\
\lambda_{2}/2\qquad & \lambda_6/2\qquad & \lambda_\xi
\end{array}\right]~,\qquad\quad
\mathcal{M}_2=~\left[\begin{array}{c c c}
\lambda_\phi\qquad & \lambda_4/2\qquad & \lambda_5/2\\
\lambda_4/2\qquad  & \lambda_\eta\qquad & \lambda_6/2\\
\lambda_5/2\qquad  & \lambda_6/2\qquad & \lambda_\xi
\end{array}\right]~,\nonumber\\
\mathcal{M}_3=&~\left[\begin{array}{c c c}
\lambda_\phi\qquad & \lambda_3/2\qquad &  \lambda_5/2\\
\lambda_3/2\qquad & \lambda_H\qquad  & \lambda_{2}/2\\
\lambda_5/2\qquad & \lambda_{2}/2\qquad & \lambda_\xi
\end{array}\right]~,\qquad\quad
\mathcal{M}_4=~\left[\begin{array}{c c c c}
\lambda_\phi\qquad & \lambda_3/2\qquad & \lambda_4/2\\
\lambda_3/2\qquad & \lambda_H\qquad & \lambda_1/2\\
\lambda_4/2\qquad & \lambda_1/2\qquad & \lambda_\eta
\end{array}\right]~.
\end{align}
In the notation $\mathcal{M}_k$, the subscript $k$~[$k=1,\,2,\,3,\,4$] denotes the index of the deleted row and column of $\Omega$. Following the CHL theorem, to ensure the strict copositivity of $\Omega$, all four principal submatrices have to be strictly copositive and at least det$(\Omega)>0$. Thus, considering the copositivity conditions for a symmetric $3\times 3$ matrix~\cite{HADELER198379, CHANG1994113}, one can write,
\begin{itemize}
\item \underline{$\mathcal{M}_1$ is strictly copositive iff :}
\begin{align}
&\lambda_H>0,~~\lambda_\eta>0,~~ \lambda_\xi>0\,,\nonumber\\
&\bar{\lambda}_1=\lambda_1+2\sqrt{\lambda_H\lambda_\eta}~~>0\,,\nonumber\\
&\bar{\lambda}_6=\lambda_6+2\sqrt{\lambda_\eta\lambda_\xi}~~>0\,,\nonumber\\
&\bar{\lambda}_{2}=\lambda_{2}+2\sqrt{\lambda_H\lambda_\xi}~~>0\,,\nonumber\\
&\lambda_1\sqrt{\lambda_\xi}+\lambda_{2}\sqrt{\lambda_\eta}+\lambda_6\sqrt{\lambda_H}+2\sqrt{\lambda_H\lambda_\eta\lambda_\xi}+\sqrt{\bar{\lambda}_1\,\bar{\lambda}_6\,\bar{\lambda}_{2}}~~>0~.
\label{eq:m1}
\end{align}
\item \underline{$\mathcal{M}_2$ is strictly copositive iff :}
\begin{align}
&\lambda_\phi>0,~~\lambda_\eta>0,~~ \lambda_\xi>0\,,\nonumber\\
&\bar{\lambda}_4=\lambda_4+2\sqrt{\lambda_\phi\lambda_\eta}~~>0\,,\nonumber\\
&\bar{\lambda}_6=\lambda_6+2\sqrt{\lambda_\eta\lambda_\xi}~~>0\,,\nonumber\\
&\bar{\lambda}_5=\lambda_5+2\sqrt{\lambda_\phi\lambda_\xi}~~>0\,,\nonumber\\
&\lambda_4\sqrt{\lambda_\xi}+\lambda_5\sqrt{\lambda_\eta}+\lambda_6\sqrt{\lambda_\phi}+2\sqrt{\lambda_\phi\lambda_\eta\lambda_\xi}+\sqrt{\bar{\lambda}_4\,\bar{\lambda}_6\,\bar{\lambda}_5}~~>0~.
\label{eq:m2}
\end{align}
\item \underline{$\mathcal{M}_3$ is strictly copositive iff :}
\begin{align}
&\lambda_\phi>0,~~\lambda_H>0,~~ \lambda_\xi>0\,,\nonumber\\
&\bar{\lambda}_3=\lambda_3+2\sqrt{\lambda_\phi\lambda_H}~~>0\,,\nonumber\\
&\bar{\lambda}_{2}=\lambda_2+2\sqrt{\lambda_H\lambda_\xi}~~>0\,,\nonumber\\
&\bar{\lambda}_5=\lambda_5+2\sqrt{\lambda_\phi\lambda_\xi}~~>0\,,\nonumber\\
&\lambda_3\sqrt{\lambda_\xi}+\lambda_5\sqrt{\lambda_H}+\lambda_{2}\sqrt{\lambda_\phi}+2\sqrt{\lambda_\phi\lambda_H\lambda_\xi}+\sqrt{\bar{\lambda}_3\,\bar{\lambda}_{2}\,\bar{\lambda}_5}~~>0~.
\label{eq:m3}
\end{align}
\item \underline{$\mathcal{M}_4$ is strictly copositive iff :}
\begin{align}
&\lambda_\phi>0,~~\lambda_H>0,~~ \lambda_\eta>0\,,\nonumber\\
&\bar{\lambda}_3=\lambda_3+2\sqrt{\lambda_\phi\lambda_H}~~>0\,,\nonumber\\
&\bar{\lambda}_4=\lambda_4+2\sqrt{\lambda_\phi\lambda_\eta}~~>0\,,\nonumber\\
&\bar{\lambda}_1=\lambda_1+2\sqrt{\lambda_H\lambda_\eta}~~>0\,,\nonumber\\
&\lambda_3\sqrt{\lambda_\eta}+\lambda_4\sqrt{\lambda_H}+\lambda_1\sqrt{\lambda_\phi}+2\sqrt{\lambda_\phi\lambda_H\lambda_\eta}+\sqrt{\bar{\lambda}_3\,\bar{\lambda}_4\,\bar{\lambda}_1}~~>0~.
\label{eq:m4}
\end{align}
\end{itemize}
Therefore, combining Eqs.~\eqref{eq:m1}$-$\eqref{eq:m4} and the condition det$(\Omega)>0$, the vacuum stability conditions for the proposed scalar configuration can be cast as,
\begin{align}
\bullet~~~~ &\lambda_\phi>0,~~\lambda_H>0,~~\lambda_\eta>0,~~ \lambda_\xi>0\,,\nonumber\\
\bullet~~~~ &\bar{\lambda}_1=\lambda_1+2\sqrt{\lambda_H\lambda_\eta}~~>0\,,\nonumber\\
\bullet~~~~ &\bar{\lambda}_{2}=\lambda_{2}+2\sqrt{\lambda_H\lambda_\xi}~~>0\,,\nonumber\\
\bullet~~~~ &\bar{\lambda}_3=\lambda_3+2\sqrt{\lambda_\phi\lambda_H}~~>0\,,\nonumber\\
\bullet~~~~ &\bar{\lambda}_4=\lambda_4+2\sqrt{\lambda_\phi\lambda_\eta}~~>0\,,\nonumber\\
\bullet~~~~ &\bar{\lambda}_5=\lambda_5+2\sqrt{\lambda_\phi\lambda_\xi}~~>0\,,\nonumber\\
\bullet~~~~ &\bar{\lambda}_6=\lambda_6+2\sqrt{\lambda_\eta\lambda_\xi}~~>0\,,\nonumber\\
\bullet~~~~ &\lambda_1\sqrt{\lambda_\xi}+\lambda_{2}\sqrt{\lambda_\eta}+\lambda_6\sqrt{\lambda_H}+2\sqrt{\lambda_H\lambda_\eta\lambda_\xi}+\sqrt{\bar{\lambda}_1\,\bar{\lambda}_6\,\bar{\lambda}_{2}}~~>0\,,\nonumber\\
\bullet~~~~ &\lambda_4\sqrt{\lambda_\xi}+\lambda_5\sqrt{\lambda_\eta}+\lambda_6\sqrt{\lambda_\phi}+2\sqrt{\lambda_\phi\lambda_\eta\lambda_\xi}+\sqrt{\bar{\lambda}_4\,\bar{\lambda}_6\,\bar{\lambda}_5}~~>0\,,\nonumber\\
\bullet~~~~ &\lambda_3\sqrt{\lambda_\xi}+\lambda_5\sqrt{\lambda_H}+\lambda_{2}\sqrt{\lambda_\phi}+2\sqrt{\lambda_\phi\lambda_H\lambda_\xi}+\sqrt{\bar{\lambda}_3\,\bar{\lambda}_{2}\,\bar{\lambda}_5}~~>0\,,\nonumber\\
\bullet~~~~ &\lambda_3\sqrt{\lambda_\eta}+\lambda_4\sqrt{\lambda_H}+\lambda_1\sqrt{\lambda_\phi}+2\sqrt{\lambda_\phi\lambda_H\lambda_\eta}+\sqrt{\bar{\lambda}_3\,\bar{\lambda}_4\,\bar{\lambda}_1}~~>0\,,\nonumber\\
\bullet~~~~ &16\lambda_\phi\lambda_H\lambda_\eta\lambda_\xi+2\lambda_1\lambda_3\lambda_4\lambda_\xi+\lambda_6^2(\lambda_3^2-4\lambda_\phi\lambda_H)+\lambda_1^2(\lambda_5^2-4\lambda_\phi\lambda_\xi)\nonumber\\
&+\lambda_2^2(\lambda_4^2-4\lambda_\phi\lambda_\eta)- 2\lambda_1\lambda_2(\lambda_4\lambda_5-2\lambda_\phi \lambda_6)+2(\lambda_5\lambda_6- \lambda_4\lambda_\xi)(2\lambda_4\lambda_H-\lambda_1\lambda_3)\nonumber\\
&-2\lambda_2\lambda_3(\lambda_4\lambda_6 - 2\lambda_5\lambda_\eta)- 4\lambda_\eta(\lambda_3^2\lambda_\xi+\lambda_5^2\lambda_H)~>0~.
\label{eq:vac}
\end{align} 
The last condition of Eq.~\eqref{eq:vac} corresponds to det$(\Omega)>0$. Note that with the real components of the scalar fields, one would get, besides Eq.~\eqref{eq:vac}, a few redundant inequalities due to the existing gauge symmetries, adding no extra constraint to the vacuum stability~\cite{Kannike:2012pe}.

After SSB, the term $\lambda_3(H^\dagger H)(\phi^\dagger \phi)$ induces a mixing between the SM Higgs and $\phi$. However, the Higgs sector is strongly constrained through various collider searches~\cite{CMS:2018amk, ATLAS:2018sbw}, resulting in an upper bound on the mixing angle~\cite{Robens:2022cun}. Therefore, one can assume $\lambda_3\to 0$ so that the presence of $\phi$ doesn't affect the Higgs observables. The vacuum stability conditions can easily be reformulated in accordance with this assumption. After the spontaneous breaking of the EW and $U(1)_{L_\mu-L_\tau}$ symmetries, the scalar sector can be redefined as,
\begin{align}
H = 
\frac{1}{\sqrt{2}}\begin{pmatrix}
0 \\
h+v
\end{pmatrix}, \quad \phi=\frac{1}{\sqrt{2}}(\phi+v^\prime), \quad \eta^-=\eta^-, \quad \xi=\xi~,
\end{align}
with $v=246$ GeV being the EW VEV. In the broken phase of $U(1)_{L_\mu-L_\tau}$, $Z^\prime$  acquires a mass $M_{Z^\prime}=g^\prime |Q_\phi| v^\prime$. Further, the physical mass terms for the scalars can be defined as,
\begin{align}
M^2_h=&~2\lambda_Hv^2~,\nonumber\\
M^2_\phi=&~2\lambda_\phi(v^\prime)^2~,\nonumber\\
M^2_{\eta}=&~\tilde{M}_{\eta}^2+\frac{\lambda_1}{2}v^2+\frac{\lambda_4}{2}(v^\prime)^2~,\nonumber\\
M^2_{\xi}=&~\tilde{M}_{\xi}^2+\lambda_2v^2+\lambda_5(v^\prime)^2~.
\label{eq:mass}
\end{align}
Here, $M_h=125$ GeV stands for the SM Higgs mass. Though with a vanishing $\lambda_3$ the decay processes $h\to\phi\phi$ or $\phi \to hh$ are absent at the tree-level, one can assume $M_h/2<M_\phi<2M_h$ to forbid any loop-induced BSM correction to the decay or production of the SM Higgs in the presence of $\phi$.
\subsection{Collider Constraints}
\label{sec:3col}
The proposed BM-$U(1)_{L_\mu-L_\tau}$ formulation contains a singly charged leptophilic scalar $\eta^-$. Such BSM scalars have been extensively searched in the colliders, leading to strong exclusion limits on the associated parameter space. However, the bounds are specific to the assumed decay channels of $\eta^-$. A recent study with a lepton flavor universal $\eta^-$, considering BR$(e)=$\,\,BR$(\mu)=25\%$ and BR$(\tau)=50\%$ has excluded $125~{\rm GeV}\leq M_\eta\leq 185$ GeV and $M_\eta\leq 80$ GeV at $95\%$ confidence level~(CL)~\cite{Das:2025oww}. Here, the notation BR$(\ell)$ denotes the branching ratio of $\eta^-$ to the SM lepton $\ell$ and an SM-singlet fermion protected by some discrete symmetry~(e.g., $Z_2$ or $R$-parity). However, the benchmark scenario can't be considered for the present model as $\eta^-$ exhibits only electron-specific Yukawa interaction, i.e., BR$(e)=100\%$. Thus, in the BM-$U(1)_{L_\mu-L_\tau}$, $\eta^-$ is phenomenologically equivalent to the selectron~($\tilde{e}$) while $\chi_2$ can be mapped into the neutralino. At the colliders, $\eta^-$ can be produced through the $s$-channel exchange of a $Z$ boson or a virtual photon. Therefore, in a hadronic\,(leptonic) collider $pp\,(\ell\ell)\to \eta^+\eta^-\to e^+e^-+\slashed{E}_T$ leads to the most prominent production signal of $\eta^-$. The ATLAS search with an integrated luminosity~(IL) of $139$ fb$^{-1}$ at $\sqrt{s}=13$ TeV for $\tilde{e}_R$~($\equiv \eta^-$) has excluded the mass range $[120,\,425]$ GeV at $95\%$ CL~\cite{ATLAS:2019lff}. However, a more stringent lower bound on $\tilde{e}_R$ can be obtained from Ref.~\cite{ATLAS:2014zve}. Thus, combining the ATLAS results from Refs.~\cite{ATLAS:2014zve, ATLAS:2019lff}, $\eta^-$ can be excluded within $[107,\,425]$ GeV at $95\%$ CL. Further, the Large Electron-Positron~(LEP) collider searches for $\tilde{e}_R$ with IL of $9.4$ pb$^{-1}$ at an average center of mass energy of 208 GeV have excluded masses below 100 GeV~\cite{ALEPH:2001oot}. Therefore, for the singly charged scalar $\eta^-$ with BR$(\eta^-\to e^-\chi_2)=100\%$, the allowed mass regime can be defined as $M_\eta\in[100,\,107]$ GeV and $M_\eta>425$ GeV.

One should note that, to satisfy Eq.~\eqref{cancel2} with $M_\eta\geq 100$ GeV, a TeV-scale $\psi$ is required in the theory. Thus, the considered phenomenology demands a model-dependent bound that allows $\psi$ for $m_\psi\in [1.68,\,1.79]$ TeV or $m_\psi>7.14$ TeV. It is worth mentioning that the considered mass regime of $\psi$ is in good agreement with the current~\cite{L3:2001xsz, ATLAS:2015qoy} and future~\cite{Shang:2021mgn} collider searches for a charged VLL. The production of the real SM-singlet scalar $\xi$ depends on its coupling with the SM Higgs\,\footnote{SM-singlet scalars that don't couple to the SM Higgs or a doublet state, can also be produced at the colliders through loop-induced couplings~\cite{Bhaskar:2020kdr,De:2024tbo}.} and can't be constrained through the colliders. In general, $M_\xi$ can be much lighter than the EW scale. However, the present analysis will be confined to a heavier mass regime where $M_\xi\geq 100$ GeV.
\subsection{Existing Bounds on the Minimal $\mathbf{\mathcal{G}_{\rm SM}\otimes U(1)_{L_\mu-L_\tau}}$}
\label{subsec:2C}
In the physical basis, the BSM contributions from the minimal gauged $U(1)_{L_\mu-L_\tau}$ theory are represented by the 2-dimensional parameter space, $M_{Z^\prime}-g^\prime$. However, experiments searching for a neutral leptophilic gauge boson have already probed a significant portion of the parameter space, leading to stringent exclusion limits. For example, the dimuon production due to the scattering of $\nu_\mu$ in the Coulomb potential of a nucleus introduces a strong constraint over the entire mass range of $Z^\prime$~\cite{Altmannshofer:2014pba, Altmannshofer:2016jzy}. A combined result from the CHARM-II~\cite{CHARM-II:1990dvf}, CCFR~\cite{CCFR:1991lpl} and NuTeV~\cite{NuTeV:1998khj} for the neutrino trident production cross section has exclued $\left(g^\prime/M_{Z^\prime}\right)^2> 3.61\times 10^{-6}~{\rm GeV}^{-2}$~\cite{Altmannshofer:2016jzy}. Further, a recent update from NA64$\mu$ has excluded the parameter space for $\{M_{Z^\prime}\leq 0.1 ~{\rm GeV},\, g^\prime\geq 6\times 10^{-4}\}$~\cite{NA64:2024klw}. 
\begin{figure}[!ht]
\centering
\includegraphics[scale=1]{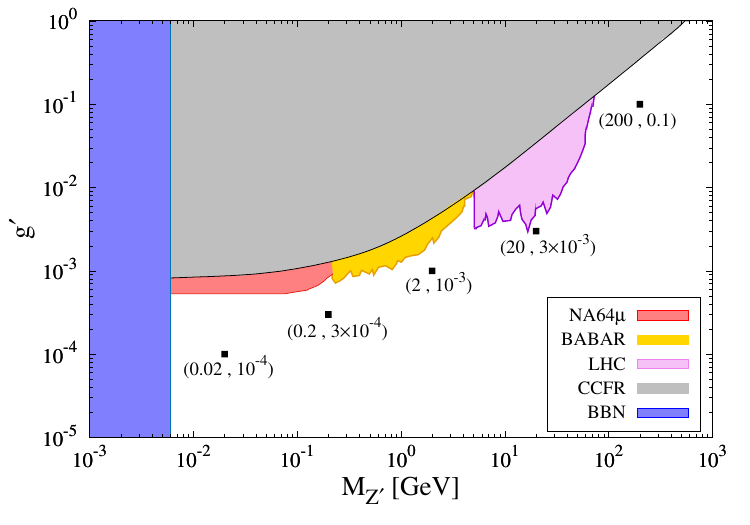}
\caption{Current constraints on the $U(1)_{L_\mu-L_\tau}$ theory. The grey, blue, red, golden, and violet shaded regions represent the parts of the parameter space excluded through the CCFR~\cite{CCFR:1991lpl}, BBN~\cite{Escudero:2019gzq}, NA64$\mu$~\cite{NA64:2024klw}, BABAR~\cite{BaBar:2016sci}, and LHC~\cite{ATLAS:2023vxg} results, respectively.}
\label{fig:U1}
\end{figure}
ATLAS and CMS searches for $Z^\prime$ production from the final state radiation of $\mu$ or $\tau$ leptons in the Drell-Yan process introduce the strongest bounds on $U(1)_{L_\mu-L_\tau}$ for $M_{Z^\prime}\in[5,\,81]$ GeV~\cite{CMS:2018yxg, ATLAS:2023vxg}, whereas a less stringent bound can be derived on $Z^\prime$ from the LEP-1 measurements for 4-fermion final states at the $Z$ pole~\cite{ALEPH:1994bie}. For $0.2~{\rm GeV}<M_{Z^\prime}\leq 10$ GeV, a tight exclusion limit can be obtained from the $e^+e^-\to \mu^+\mu^-Z^\prime\to 4\mu$ searches at the BABAR~\cite{BaBar:2016sci}. In the light $Z^\prime$ regime, the most stringent constraint comes from the big-bang nucleosynthesis~(BBN) excluding a major part of the mass range $1~{\rm eV}\leq M_{Z^\prime}< 6$ MeV for $g^\prime>10^{-13}$~\cite{Escudero:2019gzq}. Moreover, for $M_{Z^\prime}\leq 0.1$ MeV, the measurements from the stellar cooling are crucial to set bounds on the parameter space~\cite{An:2013yfc,Hardy:2016kme}. For a more detailed discussion on the recent constraints on $U(1)_{L_\mu-L_\tau}$, the reader may refer to the Refs.~\cite{Bauer:2018onh, Dasgupta:2023zrh}.

Fig.~\ref{fig:U1} displays a compilation of the strongest experimental bounds on the parameter space $\{1~{\rm MeV}\leq M_{Z^\prime}\leq 1~{\rm TeV},\, 10^{-5}\leq g^\prime\leq 1\}$ for the minimal $U(1)_{L_\mu-L_\tau}$ model. The exclusion limits from the BBN, neutrino trident production, NA64$\mu$, $4\mu$ signals at the BABAR and LHC have been depicted with blue, grey, red, golden, and violet, respectively. Further, five different points have been marked in the allowed region of Fig.~\ref{fig:U1}, particularly chosen to represent different mass regimes of $Z^\prime$. The points being consistent with the current constraints on the $U(1)_{L_\mu-L_\tau}$, can be used to explore the DM phenomenology in the next section.   
\section{Dark Matter Phenomenology}
\noindent
\label{sec:4}
As mentioned in Sec.~\ref{sec:3}, the BM-$U(1)_{L_\mu-L_\tau}$ model can explain the observed DM abundance if the SM-singlet VLL $\chi_1$ were considered as a viable DM candidate. The absolute stability of $\chi_1$ is ensured by its non-trivial $U(1)_{L_\mu-L_\tau}$ charge  and the unbroken $Z_2$ symmetry. Moreover, an assumed mass hierarchy, $m_\psi>M_{\xi}\geq M_{\eta}>m_2>m_1$ kinematically stabilizes $\chi_1$ among the $Z_2$-odd states after SSB. Note that, though $\chi_2$ is also an SM-singlet VLL stabilized through the same $Z_2$ symmetry, the considered $U(1)_{L_\mu-L_\tau}$ charge assignments prohibit the possibility of coannihilation in the present model. Further, $\chi_1$ and $\chi_2$ are assumed to be notably non-degenerate such that $(m_2-m_1)/m_1>0.1$~\cite{Mizuta:1992qp}. To be specific, $m_2=2m_1$ has been considered for the analysis. Thus, in the proposed formulation, the DM relic abundance can be fully explained through the annihilation of $\chi_1$ to the SM particles, with $Z^\prime$ being the only interaction portal. From Eq.~\eqref{eq:NP}, the $\bar{\chi}_1\chi_1 Z^\prime$ interaction can be cast as,
\begin{align}
\mathcal{L}_{\rm int}^{\chi_1}=g^\prime Q_1\bar{\chi}_1\gamma^\mu\chi_1 Z^\prime_\mu=\rho_1\bar{\chi}_1\gamma^\mu\chi_1 Z^\prime_\mu~,
\end{align}
where, $\rho_1=g^\prime Q_1$. From the theoretical perspective, $Q_1$ is a free parameter, and in principle, can assume any arbitrary value as long as the perturbative unitarity is maintained for $\rho_1$, i.e., $|\rho_1|<4\pi$~\cite{Allwicher:2021rtd}. Therefore, as a moderate choice, $Q_1$ can be considered 10 so that for $g^\prime\leq 1$, $\rho_1$ doesn't violate the PU bound. However, in the literature, much larger values of $Q_1$ have also been used to explore the DM phenomenology~\cite{Barman:2024lxy,De:2024tvj}. The total annihilation cross section of $\chi_1$ being significantly dependent on the chosen $U(1)_{L_\mu-L_\tau}$ gauge coupling and the $Z^\prime$ mass, one must consider only those regions of the allowed $M_{Z^\prime}-g^\prime$ parameter space where the present relic abundance can be explained. Assuming $\chi_1$ to be a cold dark matter~(CDM), the observed Planck data~\cite{Planck:2018vyg} can be addressed either through the Freeze-out or Freeze-in mechanism, depending on the effective coupling strength between the DM and the SM sector. However, at the scale of feeble interaction~(required for Freeze-in), i.e., $g^\prime\sim\mathcal{O}(10^{-11})$, the minimal $\mathcal{G}_{\rm SM}\otimes U(1)_{L_\mu-L_\tau}$ model contributes negligibly to the considered leptonic observables and becomes effectively redundant in the BM-$U(1)_{L_\mu-L_\tau}$ setup. Therefore, the DM phenomenology will be broadly restricted to $\{M_{Z^\prime}\geq 10~{\rm MeV},\,g^\prime\geq 10^{-4}\}$ --- a parameter space that can be probed through the future experiments, e.g., NA64$\mu$~\cite{Sieber:2021fue}, SHiP~\cite{Alekhin:2015byh}, FCC-ee~\cite{Airen:2024iiy}, and M$\mu$C~\cite{Huang:2021nkl}. Clearly, with the assumed interaction strength, the Freeze-out mechanism is the only natural choice.  
\begin{figure}[!ht]
\centering
\includegraphics[scale=0.62]{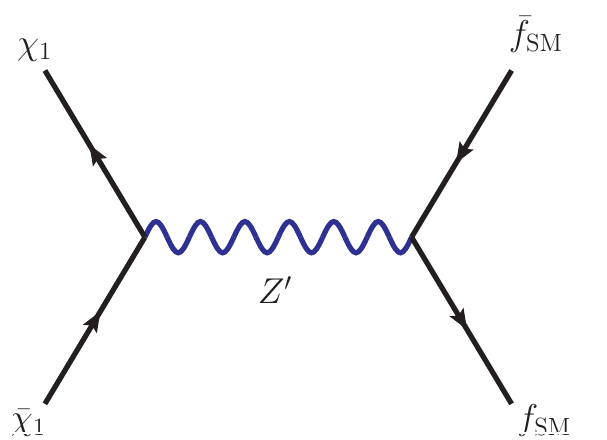}
\caption{Possible $s$-channel diagrams contributing to the total annihilation cross section of $\chi_1$. For $m_1>m_\tau$, $f_{\rm SM}\equiv\mu$, $\nu_\mu$, $\tau$, $\nu_\tau$.}
\label{fig:anni}
\end{figure}

Fig.~\ref{fig:anni} is a representative diagram for all the $s$-channel annihilation processes contributing to the relic density computation. $Z^\prime$ being a leptophilic neutral gauge boson with flavor-specific interactions, $f_{\rm SM}$ can be $\mu$, $\nu_\mu$, $\tau$, $\nu_\tau$ as long as $m_1>m_\tau$. However, for $m_1\leq m_\mu$, $\chi_1$ can only annihilate to the $2^{\rm nd}$ and $3^{\rm rd}$ generation neutrinos.
\subsection{Relic Density}
In the early universe, the thermal equilibrium between $\chi_1$ and the SM sector particles~($f_{\rm SM}$) was ensured by the fact that $10^{-3}\leq \rho_1\leq 1$. However, at a later time, when the expansion rate of the universe exceeded the interaction rate between $\chi_1$ and $f_{\rm SM}$, $\chi_1$ decoupled from the thermal bath with its abundance being frozen forever to the decoupling value, i.e., the relic density $\Omega_1 h^2$. For a single-component DM theory, the relic abundance can be obtained by solving the Boltzmann equation: 
\begin{align}
\frac{dn_1}{dt}+3\mathcal{H}n_1=-\left\langle\sigma_{\rm An}|\mathbf{v}|\right\rangle\Big[n_1^2-(n_1^{\rm eq})^2\Big]~,
\label{eq:boltz}
\end{align}
where $n_1$ stands for the number density of $\chi_1$ with the superscript `eq' representing its equilibrium value. $\mathcal{H}$ defines the Hubble parameter, and $\left\langle\sigma_{\rm An}|\mathbf{v}|\right\rangle$ is the thermal averaged annihilation cross section times the relative velocity~($\mathbf{v}$) of $\chi_1$.
\begin{figure}[!ht]
\centering
\includegraphics[scale=1]{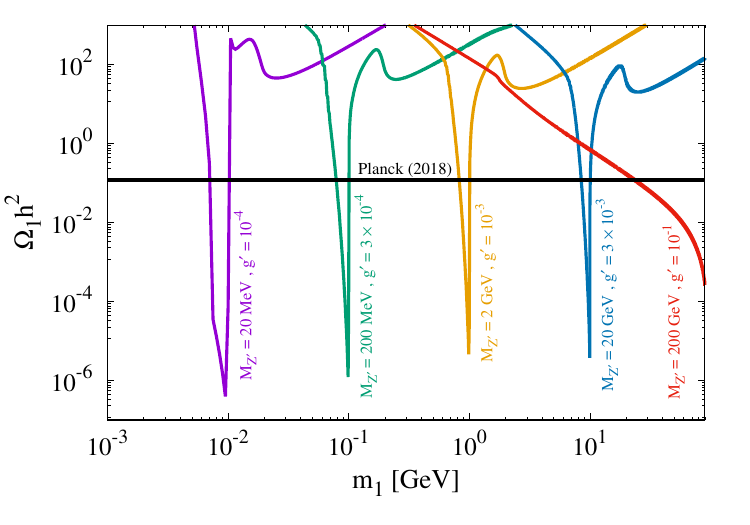}
\caption{Variation of $\Omega_1 h^2$ as a function of $m_1$. The colors violet, green, golden, blue, and red correspond to $\left(M_{Z^\prime}=20~{\rm MeV},\, g^\prime=10^{-4}\right)$, $\left(M_{Z^\prime}=200~{\rm MeV},\, g^\prime=3\times 10^{-4}\right)$, $\left(M_{Z^\prime}=2~{\rm GeV},\, g^\prime=10^{-3}\right)$, $\left(M_{Z^\prime}=20~{\rm GeV},\, g^\prime=3\times 10^{-3}\right)$, and $\left(M_{Z^\prime}=200~{\rm GeV},\, g^\prime=0.1\right)$, respectively. For a given $g^\prime$, $\rho_1=10g^\prime$. The black line represents the observed DM abundance from the Planck~\cite{Planck:2018vyg}.}
\label{fig:relic}
\end{figure}
Note that within the assumed gauge extension, the DM observables must be in agreement with the experimentally allowed $M_{Z^\prime}- g^\prime$ parameter space. Thus, as stated earlier, the set of five distinct $\left(M_{Z^\prime},\, g^\prime\right)$ values from Fig.~\ref{fig:U1} has been considered as the NP input to test if $\Omega_1h^2=0.1198\pm 0.0012$~\cite{Planck:2018vyg} within the range $m_1\in[1~{\rm MeV},\, 90~{\rm GeV}]$. For the present work, Eq.~\eqref{eq:boltz} has been solved numerically using $\mathtt{micrOMEGAs}$~\cite{Belanger:2010pz}, while the corresponding model files have been generated through the $\mathtt{LanHEP}$~\cite{Semenov:2008jy}. 
Fig.~\ref{fig:relic} shows the variation of $\Omega_{1}h^2$ as a function of $m_1$. The horizontal black line defines the central value of the observed DM abundance. In the plot, violet, green, golden, blue, and red have been used to label $\left(M_{Z^\prime}=20~{\rm MeV},\, g^\prime=10^{-4}\right)$, $\left(M_{Z^\prime}=200~{\rm MeV},\, g^\prime=3\times 10^{-4}\right)$, $\left(M_{Z^\prime}=2~{\rm GeV},\, g^\prime=10^{-3}\right)$, $\left(M_{Z^\prime}=20~{\rm GeV},\, g^\prime=3\times 10^{-3}\right)$, and $\left(M_{Z^\prime}=200~{\rm GeV},\, g^\prime=0.1\right)$, respectively. One can readily visualize the positions of these points on the $M_{Z^\prime}-g^\prime$ plane in Fig.~\ref{fig:U1}. In each of the cases, $\chi_1$ is found to satisfy the observed relic density near the resonance funnel. Thus, one can tabulate a few benchmark points~(BP) that are simultaneously consistent with the DM phenomenology and the searches for a hidden gauge sector. Check Table~\ref{tab:DM} for the list.
\begin{table}[!ht]
\centering
\begin{tabular}{|c|c|c|c|c|c|}
\hline
Benchmark & $\qquad M_{Z^\prime}\qquad$ & $\qquad\quad g^\prime\qquad\quad$ & $\qquad\qquad m_1\qquad\qquad$ & $\qquad\Omega_1 h^2\quad$ & $\qquad\qquad m_2\qquad\qquad$\\
Points & [GeV] & & [GeV] & & [GeV]\\
\hline\hline
BP1 & $0.02$ & $1\times10^{-4}$ & $7.27\times 10^{-3}$ &  $0.121$ & $14.54\times 10^{-3}$\\
\hline
BP2 & $0.2$ & $3\times10^{-4}$ & $78.85\times 10^{-3}$ & $0.120$ & $15.77\times 10^{-2}$\\
\hline
BP3 & $2$ & $1\times10^{-3}$ & $8.26\times 10^{-1}$ & $0.120$ & $16.52\times 10^{-1}$\\
\hline
BP4 & $20$ & $3\times 10^{-3}$ & $8.49$ & $0.119$ & $16.98$\\
\hline
BP5 & $200$ & $1\times10^{-1}$ & $23.40$ & $0.121$ & $46.80$\\
\hline
\end{tabular}
\caption{The list of parameter space points where the present DM abundance and the experimental constraints on a $U(1)_{L_\mu-L_\tau}$ gauge extension can be simultaneously satisfied.}
\label{tab:DM}
\end{table}

Note that apart from the perturbative unitarity, $\rho_1$ can also be constrained through the ellipticity of the galactic DM halos. The resultant upper limit can be read as~\cite{Feng:2009hw,Essig:2011nj},
\begin{align}
\rho_1\leq 0.1\left(\frac{M_{Z^\prime}}{10~{\rm MeV}}\right)\left(\frac{m_1}{100~{\rm MeV}}\right)^{-1/4}~.
\end{align}
One can easily verify that all five benchmark points of Table~\ref{tab:DM} are compatible with this bound. 
\subsection{Direct Detection Prospects}
The conventional modes of DD experiments fail to trace the DM-SM interactions within the proposed BM-$U(1)_{L_\mu-L_\tau}$ model, as both of the $\chi_1$-electron and $\chi_1$-quark scattering amplitudes are proportional to the kinetic mixing. The NP fields $\eta$ and $\psi$ being correlated through Eq.~\eqref{cancel2}, the leading order kinetic mixing vanishes in the present theory and it becomes trivial to explain the null results in the existing DD searches. However, in a very recent proposal from Peking University~(PKU), China, a new direct detection strategy has been introduced~\cite{Ruzi:2023mxp, Yu:2024spj} which might be significant for the present framework. The proposed experiment is currently planning to use cosmic muons and, in a future update, high-energy muon beams to scatter on the DM. For the ultra-relativistic muon beams, the DM can be considered as a quasi-static target, effectively frozen within the detector. The deflection of the incident muon beams through a vacuum can be detected by the multiple layers of a Gas Electron Multiplier~(GEM), indicating the presence of a $\mu$-philic DM. However, the scattering cross section significantly depends on the center-of-mass energy of the DM-muon 2-body system. Therefore, the detection prospects can be parametrized as,
\begin{align}
\sigma_{\chi_1\mu}(q^2)=\Theta(q^2)\bar{\sigma}_{\chi_1\mu}~,
\end{align}
where, 
\begin{align}
\bar{\sigma}_{\chi_1\mu}=\frac{(g^\prime)^4Q_1^2}{\pi\,M^4_{Z^\prime}}\left(\frac{m_1m_\mu}{m_1+m_\mu}\right)^2
\end{align}
defines the DM-muon elastic scattering cross section in the non-relativistic limit. $\Theta(q^2)$ is the relativistic correction that tends to 1 as $q^2\to 0$. Here, $q$ denotes the transferred momentum. One can easily calculate the explicit form of $\Theta(q^2)$ from the relativistic kinematics of 2-fermion scattering. However, for the present purpose, it's sufficient to compute $\bar{\sigma}_{\chi_1\mu}$ --- the relativistic correction just adds an extra suppression.
\begin{figure}[!ht]
\centering
\includegraphics[scale=1]{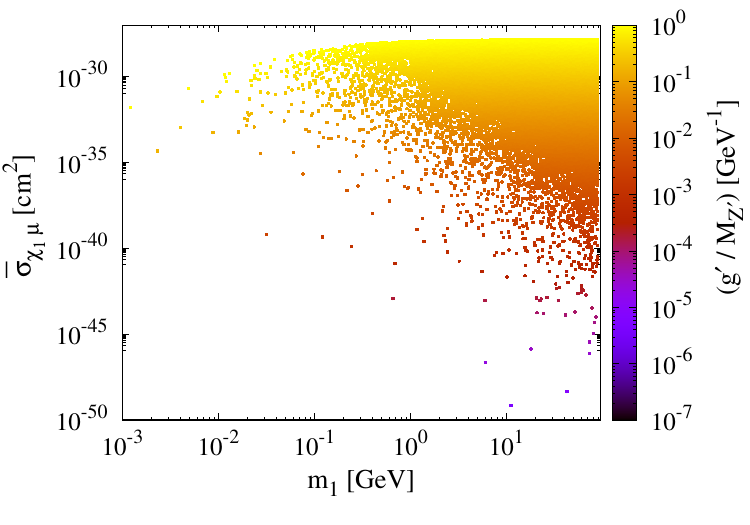}
\caption{Variation of $\bar{\sigma}_{\chi_1\mu}$ as a function of the DM mass $m_1$ for $(g^\prime/M_{Z^\prime})\in\,[10^{-7},\,1]$ GeV$^{-1}$.}
\label{fig:sig}
\end{figure}
Fig.~\ref{fig:sig} presents the variation of $\bar{\sigma}_{\chi_1\mu}$ as a function of $m_1$ for different values of $(g^\prime/M_{Z^\prime})$ ranging between $10^{-7}$ GeV$^{-1}$ to 1 GeV$^{-1}$. Clearly, $\bar{\sigma}_{\chi_1\mu}$, and hence $\sigma_{\chi_1\mu}(q^2)$, is far below the projected sensitivity of the current and future PKU-muon experiments~\cite{Yu:2024spj}. However, in principle, the proposal can be crucial to test/falsify the BM-$U(1)_{L_\mu-L_\tau}$ once the detector threshold is significantly improved.
\section{New Physics Contributions to \,\pmb{$Z\to\ell^+\ell^-$}}
\noindent
\label{sec:5}
The BM-$U(1)_{L_\mu-L_\tau}$ model can modify the $Z\ell^+\ell^-$~[$\ell=e,\,\mu,\,\tau$] couplings at the one-loop level, leading to significant constraints on the associated parameter space. In the minimal $U(1)_{L_\mu-L_\tau}$ theory, one-loop BSM contributions can be generated only for the $2^{\rm nd}$ and $3^{\rm rd}$ generation leptons through the $Z^\prime$ exchange. However, in the presence of the NP Yukawa couplings of Eq.~\eqref{eq:NP}, $Z\to e^+e^-$ can also be corrected, whereas an additional one-loop contribution to $Z\to \mu^+\mu^-$ can arise for $y_\mu\neq 0$. 

The partial decay width of $Z$ to the SM leptons is an experimentally well-measured observable and can be read as~\cite{ParticleDataGroup:2024cfk},
\begin{align}
\Gamma_{Z\ell\ell}=\left(83.984\pm 0.086\right)~{\rm MeV}~.
\end{align}
Thus, one can parametrize the BSM correction as $\Delta \Gamma_{Z\ell\ell}=\left|\Gamma^{\rm SM+BSM}_{Z\ell\ell}-\Gamma^{\rm SM}_{Z\ell\ell}\right|< 0.086$ MeV. 

\begin{figure}[!ht]
\centering
\subfloat[(a)]{\includegraphics[scale=0.6]{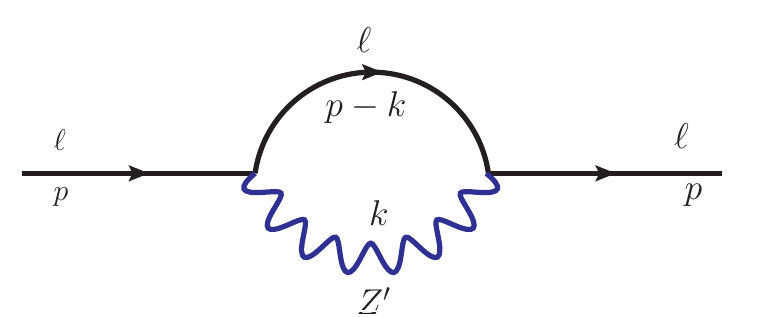}}\qquad\,
\subfloat[(b)]{\includegraphics[scale=0.6]{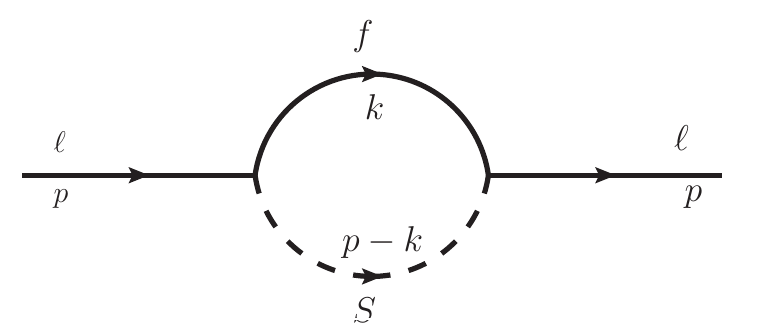}}\\
\subfloat[(c)]{\includegraphics[scale=0.58]{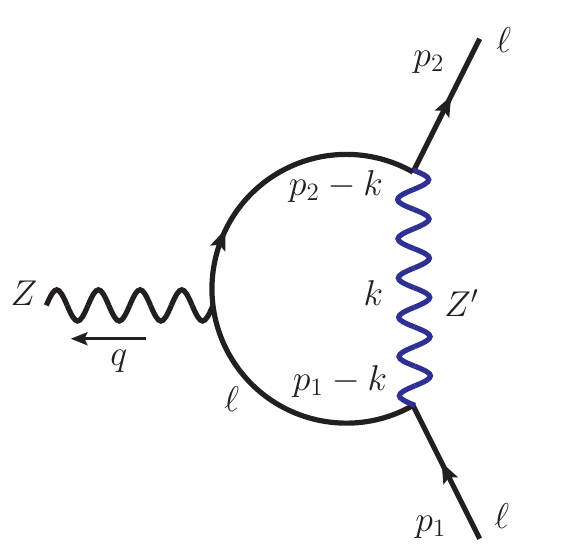}}
\subfloat[(d)]{\includegraphics[scale=0.58]{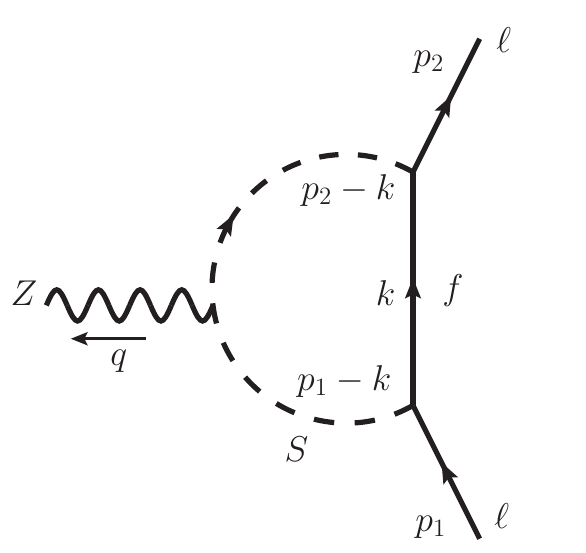}}
\subfloat[(e)]{\includegraphics[scale=0.58]{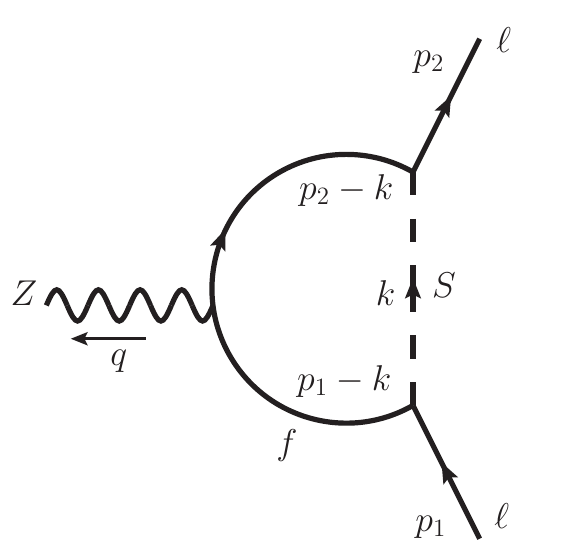}}\\
\subfloat[(f)]{\includegraphics[scale=0.6]{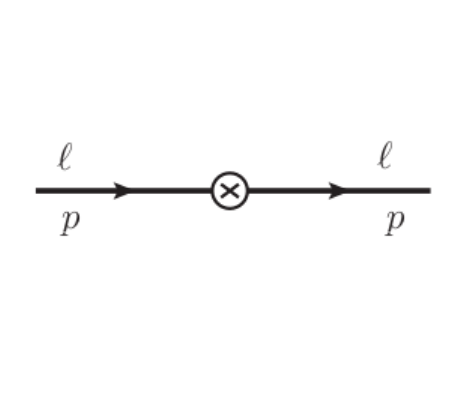}}\qquad\qquad\qquad\qquad\qquad
\subfloat[(g)]{\includegraphics[scale=0.6]{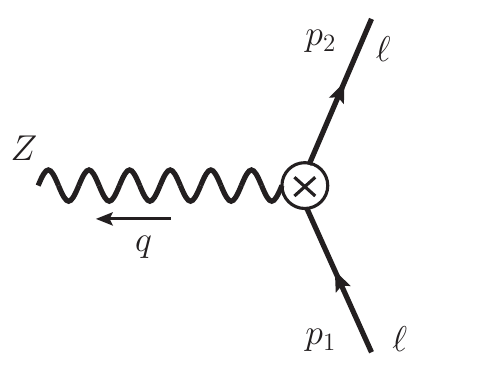}}
\caption{BSM corrections to the lepton self-energy and $Z\ell^+\ell^-$ vertex along with the respective counter terms arising through the wave function renormalization. $S$ and $f$ are the dummy notations for any $SU(2)_L$-singlet BSM scalar and fermion, respectively.}
\label{fig:Zll}
\end{figure}

Fig.~\ref{fig:Zll} shows the one-loop Feynman diagrams contributing to the $Z\ell^+\ell^-$ vertex and the self-energy of the SM leptons. Figs.~\ref{fig:Zll}\,(a) and \ref{fig:Zll}\,(c) correspond to the minimal $U(1)_{L_\mu-L_\tau}$ model with $\ell$ representing either a muon or a tau lepton. Figs.~\ref{fig:Zll}\,(b), \ref{fig:Zll}\,(d), and \ref{fig:Zll}\,(e) originate from the considered NP Yukawa interactions where $\ell$ can either be an electron or a muon. Figs.~\ref{fig:Zll}\,(f) and \ref{fig:Zll}\,(g) depict the counter terms required to cancel the UV divergences. Note that the chiral symmetry prohibits mass renormalization in the proposed framework.
 
\subsection{$\mathbf{Z\to e^+e^-}$}
In BM-$U(1)_{L_\mu-L_\tau}$, the term $y_e\bar{e}_R\eta^-\chi_2$ is the only source of NP correction to the $Ze^+e^-$ vertex. Thus, one can map $\chi_2$ to $f$ and $\eta$ to $S$ for the $e$-specific NP interaction. However, $\chi_2$ being SM-singlet, only the Fig.~\ref{fig:Zll}\,(d) is relavent for $Z\to e^+e^-$ decay. Therefore, the corresponding vertex correction term can be cast as,
\begin{align}
\bar{u}(p_2)\delta \mathcal{Y}^\mu_{Zee}u(p_1)=iy_e^2Y_\eta\, \bar{u}(p_2) P_L\int\frac{d^4k}{(2\pi)^4}\Bigg[\frac{\slashed{k}(p_1+p_2-2k)^\mu}{(k^2-m_2^2)\{(p_1-k)^2-M_\eta^2\}\{(p_2-k)^2-M_\eta^2\}}\Bigg]u(p_1)~.
\end{align}
Here $P_{L,\,R}$ are the chirality projectors. After Feynman parametrization, $\delta \mathcal{Y}^\mu_{Zee}$ can be expressed as,
\begin{align}
\bar{u}(p_2)\delta \mathcal{Y}^\mu_{Zee}u(p_1)=&~2iy_e^2Y_\eta\, \bar{u}(p_2)P_L\int^1_0dx\int^{1-x}_0dz\int\frac{d^4P}{(2\pi)^4}\Bigg[-\frac{2\slashed{P}P^\mu}{(P^2-\Delta_1)^3}\nonumber\\
&\qquad\qquad\qquad\quad+\frac{m_e(1-x)\{(1-2z)p_1^\mu+(2x+2z-1)p_2^\mu\}}{(P^2-\Delta_1)^3}\Bigg]u(p_1)~,
\end{align}
where, for an on-shell $Z$ decay, $\Delta_1=M_\eta^2\Big[1-x+x\left(m_2/M_\eta\right)^2-z(1-x-z)(M_Z/M_\eta)^2\Big]$. Consequently, the coefficient of $\gamma^\mu P_R$ can be obtained as,
\begin{align}
\delta \mathcal{W}_{Zee}&=\frac{y_e^2Y_\eta}{16\pi^2}\Bigg[\frac{\Delta_\zeta}{2}-\int^1_0dx\int^{1-x}_0dz\left\{\ln\Delta_1-\frac{2x(1-x)m_e^2}{\Delta_1}\right\}\Bigg]\nonumber\\
&\simeq\frac{y_e^2Y_\eta}{16\pi^2}\Bigg[\frac{\Delta_\zeta}{2}-\ln M_\eta-\int^1_0dx\int^{1-x}_0dz\,\ln\Big\{1-x+x\left(m_2/M_\eta\right)^2-z(1-x-z)(M_Z/M_\eta)^2\Big\}\Bigg]~.
\label{eq:Zll1}
\end{align}
Note that the term proportional to $(m_e/M_\eta)^2$ has been neglected in the above expression. The UV divergence is encapsulated in $\Delta_\zeta=\frac{1}{\zeta}-\gamma_E+\ln(4\pi)+\mathcal{O}(\zeta)$ with $\zeta\to 0$ in the 4-dimensions. $\gamma_E\approx 0.5772$ is the Euler-Mascheroni constant. 

The self-energy generated from Fig.~\ref{fig:Zll}\,(b) can be defined as,
\begin{align}
-i\,\Sigma^e(\slashed{p})&=y_e^2\int\frac{d^4k}{(2\pi)^4}\Bigg[\frac{\slashed{k}}{(k^2-m_2^2)\{(k-p)^2-M_\eta^2\}}\Bigg]\nonumber\\
&=y_e^2\,\mathcal{B}(p^2)\slashed{p}~,
\label{eq:SE}
\end{align}
where,
\begin{align}
\mathcal{B}(p^2)=\int_0^1dx\,(1-x)\Bigg[\frac{i}{16\pi^2}\left(\Delta_\zeta-\ln\Delta_2\right)\Bigg]~.
\end{align} 
The effective mass-squared term is given by $\Delta_2=xm_2^2+(1-x)M_\eta^2-x(1-x)p^2$. $\chi_2$ being lighter than $\eta$, $\mathcal{B}(p^2)$ can be computed as\,\footnote{A limiting case of Eq.~\eqref{eq:SE1} for $m_2\to 0$ can be found in Ref.~\cite{Brdar:2020nbj}.},
\begin{align}
\mathcal{B}(p^2)=\frac{i}{16\pi^2}\Bigg[\frac{\Delta_\zeta}{2}-\ln M_\eta+\mathcal{J}\left(\frac{m_2^2}{M_\eta^2}\right)+\mathcal{O}(p^2)+\cdots\Bigg]~,
\label{eq:SE1}
\end{align}
with the mass function $\mathcal{J}$ being defined as,
\begin{align}
\mathcal{J}(a)=\frac{1-4a+3a^2-2a^2\ln a}{4(1-a)^2}~.
\end{align}
The UV divergence in the electron self-energy can be canceled by renormalizing the wave function, $e_R\to (1+\delta^e_Z)^{1/2}e_R$, resulting in a counter term $i\delta^e_Z\bar{e}_R\slashed{D}e_R$. In the physical basis, it can further be split into two counter terms: $i\delta^e_Z\bar{e}_R\slashed{\partial}e_R$ and $-\delta^e_Zg_1\sin\theta_WY_{e_R}\bar{e}_R\gamma^\mu e_R Z_\mu$. The former corresponds to Fig.~\ref{fig:Zll}\,(f) and cancels the UV divergence arising from the self-energy diagram, whereas the latter is represented by Fig.~\ref{fig:Zll}\,(g) and cancels the UV divergence in Eq.~\eqref{eq:Zll1}. 

Using the on-shell renormalization scheme, one obtains,  
\begin{align}
\delta^e_Z=&~\frac{d\Sigma^e(\slashed{p})}{d\slashed{p}}\Bigg|_{\slashed{p}\,=\,m_e}\nonumber\\
\simeq &~\frac{-y_e^2}{16\pi^2}\Bigg[\frac{\Delta_\zeta}{2}-\ln M_\eta+\mathcal{J}\left(\frac{m_2^2}{M_\eta^2}\right)\Bigg]~.
\end{align}
As before, the electron mass has been neglected compared to the NP scale involved. Thus, the renormalized effective $Ze^+e^-$ interaction can be cast as, $-g_1\sin\theta_WY_{e_R}\left(1+\delta \mathcal{W}^{\mathbf{R}}_{Zee}\right)\bar{e}\gamma^\mu P_R e\, Z_\mu$. The superscript ``$\mathbf{R}$" defines the renormalized contribution, such that
\begin{align}
\delta \mathcal{W}^{\mathbf{R}}_{Zee}&=-\frac{y_e^2}{16\pi^2}\Bigg[\mathcal{J}\left(\frac{m_2^2}{M_\eta^2}\right)+\int^1_0dx\int^{1-x}_0dz\,\ln\Big\{1-x+x\left(m_2/M_\eta\right)^2-z(1-x-z)(M_Z/M_\eta)^2\Big\}\Bigg]\nonumber\\
&=-\frac{y_e^2}{16\pi^2}\Bigg[\mathcal{J}\left(\frac{m_2^2}{M_\eta^2}\right)+\int^1_0dx\,\,\, \mathcal{R}(x)\Bigg]~.
\label{eq:ZeeR}
\end{align}
The UV divergent parts in $\delta \mathcal{W}_{Zee}$ and $Y_{e_R}\delta^e_Z$ cancel out because $Y_\eta=Y_{e_R}$. In the Eq.~\eqref{eq:ZeeR}, $\mathcal{R}(x)$ is defined as,
\begin{align}
\mathcal{R}(x)=(1-x)\Big[\ln\left\{(1-x-z_+)(1-x-z_-)\right\}-2\Big]+z_+\ln\left(\frac{z_+}{x+z_+-1}\right)+z_-\ln\left(\frac{z_-}{x+z_--1}\right)~,
\label{eq:Mx}
\end{align}
where,
\begin{align}
z_\pm=\left(\frac{1-x}{2}\right)~\pm~\frac{1}{2(M_Z/M_\eta)} \Bigg[(1-x)^2(M_Z/M_\eta)^2-4\Big\{1-x+x(m_2/M_\eta)^2\Big\}\Bigg]^{1/2}~.
\end{align}
At the tree-level, the partial width for $Z\to \ell^+\ell^-$ decay can be formulated as,
\begin{align}
\Gamma_{Z\ell\ell}=\frac{G_FM_Z^3}{3\sqrt{2}\,\pi}\left(A_R^2+A_L^2\right)~,
\label{eq:zdec}
\end{align}
where $A_{R,L}$ are the form factors associated with $\gamma^\mu P_{R,L}$, respectively, and $G_F$ represents the Fermi constant. For the charged leptons, $A_R=\sin^2\theta_W$ and $A_L=-1/2+\sin^2\theta_W$. Since the $e$-specific NP interaction affects only the coupling between $e_R$ and $Z$, i.e., the $A_R$ component of Eq.~\eqref{eq:zdec}, the complete $Z\to e^+e^-$ decay width can be cast as,
\begin{align}
\Gamma_{Zee}^{\rm SM+BSM}=&~\frac{G_FM_Z^3}{3\sqrt{2}\,\pi}\Big[A_R^2\left(1+\delta \mathcal{W}^{\mathbf{R}}_{Zee}\right)\left(1+\delta \mathcal{W}^{\mathbf{R}}_{Zee}\right)^* +A_L^2\Big]\nonumber\\
=&~\frac{G_FM_Z^3}{3\sqrt{2}\,\pi}\left(A_R^2+A_L^2\right) +\frac{G_FM_Z^3}{3\sqrt{2}\,\pi}~A_R^2\times\Bigg(\left|\delta \mathcal{W}^{\mathbf{R}}_{Zee}\right|^2+2\,{\rm Re}\left[\delta \mathcal{W}^{\mathbf{R}}_{Zee}\right]\Bigg)~.
\label{eq:NPe1}
\end{align}
Note that the first term of Eq.~\eqref{eq:NPe1} defines the leading order SM contribution to $Z\to e^+e^-$ decay. Thus, including all the higher order SM contributions, one can recast Eq.~\eqref{eq:NPe1} as,
\begin{align}
&\Gamma_{Zee}^{\rm SM+BSM}=\Gamma_{Zee}^{\rm SM}+\frac{G_FM_Z^3}{3\sqrt{2}\,\pi}~A_R^2\times\Bigg(\left|\delta \mathcal{W}^{\mathbf{R}}_{Zee}\right|^2+2\,{\rm Re}\left[\delta \mathcal{W}^{\mathbf{R}}_{Zee}\right]\Bigg)\nonumber\\
\Rightarrow &~\Delta \Gamma_{Zee}=\frac{G_FM_Z^3}{3\sqrt{2}\,\pi}~A_R^2\,\Bigg|\left|\delta \mathcal{W}^{\mathbf{R}}_{Zee}\right|^2+2\,{\rm Re}\left[\delta \mathcal{W}^{\mathbf{R}}_{Zee}\right]\Bigg|~.
\label{eq:NPe}
\end{align} 
\begin{figure}[!ht]
\centering
\includegraphics[scale=1]{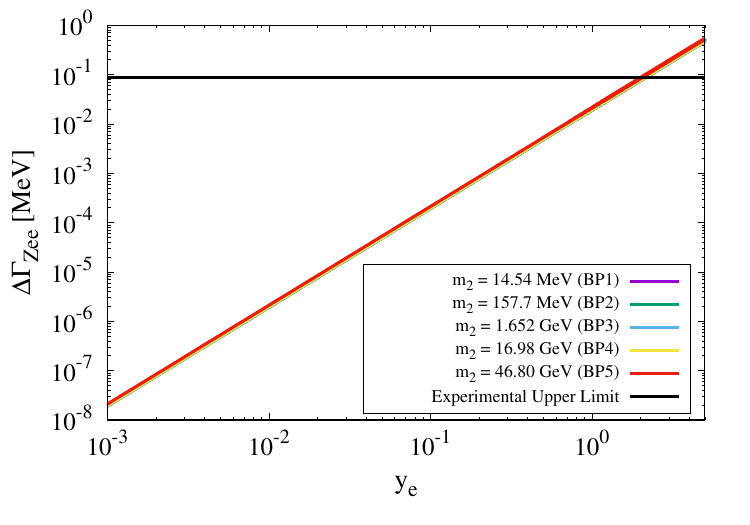}
\caption{Variation of $\Delta\Gamma_{Zee}$ as a function of the NP coupling $y_e$ for $M_\eta=100$ GeV. Different colors indicate different $m_2$ values, with the horizontal black line at 0.086 MeV representing the experimental upper limit on $\Delta\Gamma_{Zee}$.}
\label{fig:Zee}
\end{figure}
Fig.~\ref{fig:Zee} displays the variation of $\Delta\Gamma_{Zee}$ with increasing $y_e$ values. $M_\eta$ has been fixed at 100 GeV with $y_e$ varying from $10^{-3}$ to $\sqrt{8\pi}$. The choice of $M_\eta$ corresponds to the collider constraints~[discussed in Sec.~\ref{sec:3col}], whereas the $m_2$ values have been fixed through the relation $m_2=2m_1$; here the benchmark values from Table~\ref{tab:DM} have been used. However, the results are mostly independent of $m_2$ as suggested by the nearly overlapping lines. The black horizontal line marks the experimental upper bound on $\Delta\Gamma_{Zee}$, i.e., 0.086 MeV. Thus, the measurement of $Z\to e^+e^-$ decay discards $y_e> 2.01$ --- a more stringent bound than that obtained from the perturbative unitarity. 
\subsection{$\mathbf{Z\to\mu^+\mu^-}$}
\label{sec:Zmm}
The proposed model can generate two BSM correction terms to the $Z\mu^+\mu^-$ vertex at the one-loop level --- one corresponds to Fig.~\ref{fig:Zll}\,(e) arising through the NP interaction $y_\mu\bar{\mu}_R\xi\psi^-$, while the other originates from the $Z^\prime$ exchange between the muons~[see Fig.~\ref{fig:Zll}\,(c)]. The former contribution can be computed by replacing $f$ and $S$ in Fig.~\ref{fig:Zll}\,(e) with $\psi$ and $\xi$, respectively. The corresponding vertex correction is given by,
\begin{align}
\bar{u}(p_2)\delta \mathcal{Y}^\nu_{Z\mu\mu}u(p_1)=iy_\mu^2Y_\psi \, \bar{u}(p_2)P_L\int\frac{d^4k}{(2\pi)^4}\Bigg[\frac{(\slashed{p}_2-\slashed{k})\gamma^\nu(\slashed{p}_1-\slashed{k})+m_\psi^2\gamma^\nu}{(k^2-M_\xi^2)\{(p_1-k)^2-m_\psi^2\}\{(p_2-k)^2-m_\psi^2\}}\Bigg]u(p_1)\,.
\end{align}
Feynman parametrization, followed by a little rearrangement of terms, leads to 
\begin{align}
\bar{u}(p_2)\delta \mathcal{Y}^\nu_{Z\mu\mu}u(p_1)=&~2iy_\mu^2Y_\psi \, \bar{u}(p_2)P_L\int^1_0dx\int^{1-x}_0dz\int\frac{d^4P}{(2\pi)^4}\Bigg[\frac{\slashed{P}\gamma^\nu\slashed{P}}{(P^2-\Delta_3)^3}\nonumber\\
&\qquad\qquad\qquad\qquad\qquad\qquad+\frac{z(1-x-z)M_Z^2\gamma^\nu+m_\psi^2\gamma^\nu}{(P^2-\Delta_3)^3}\Bigg]u(p_1)~,
\end{align}
where, $\Delta_3=m_\psi^2\left[1-x+x(M_\xi/m_\psi)^2-z(1-x-z)(M_Z/m_\psi)^2\right]$. Defining $\delta \mathcal{Y}^\nu_{Z\mu\mu}=\delta \mathcal{W}_{Z\mu\mu}\gamma^\nu P_R$, the momentum integration results in,
\begin{align}
\delta \mathcal{W}_{Z\mu\mu}=&~\frac{y_\mu^2Y_\psi}{16\pi^2}\Bigg[\frac{\Delta_\zeta}{2}-\ln m_\psi-\int^1_0dx\int^{1-x}_0dz\Bigg\{\ln\Big[1-x+x\left(M_\xi/m_\psi\right)^2-z(1-x-z)(M_Z/m_\psi)^2\Big]\nonumber\\
&\qquad\qquad\qquad\qquad-\frac{m_\psi^2+z(1-x-z)M_Z^2}{m_\psi^2\left[1-x+x(M_\xi/m_\psi)^2-z(1-x-z)(M_Z/m_\psi)^2\right]}\Bigg\}\Bigg]~.
\label{eq:Zmm1}
\end{align}
Here, the muons have been assumed massless in comparison to $\psi$. The corresponding muon self-energy correction is generated from Fig.~\ref{fig:Zll}\,(b) and can be calculated as,
\begin{align}
-i\,\Sigma^\mu_1(\slashed{p})&=\frac{iy_\mu^2}{16\pi^2}\Bigg[\frac{\Delta_\zeta}{2}-\ln m_\psi+\mathcal{J}\left(\frac{M_\xi^2}{m_\psi^2}\right)+\mathcal{O}(p^2)+\cdots\Bigg]\slashed{p}~.
\end{align}
Further, in limit $(m_\mu/m_\psi)^2\to 0$, the counter term can be obtained as,
\begin{align}
\delta^{\mu(1)}_Z=\frac{-y_\mu^2}{16\pi^2}\Bigg[\frac{\Delta_\zeta}{2}-\ln m_\psi+\mathcal{J}\left(\frac{M_\xi^2}{m_\psi^2}\right)\Bigg]~.
\end{align}
Therefore, the renormalized contribution to $Z\mu^+\mu^-$ vertex originating from the $y_\mu\bar{\mu}_R\xi\psi^-$ interaction can be parametrized as,
\begin{align}
\delta \mathcal{W}^\mathbf{R}_{Z\mu\mu}=&~-\frac{y_\mu^2}{16\pi^2}\Bigg[\mathcal{J}\left(\frac{M^2_\xi}{m_\psi^2}\right)+\int^1_0dx\int^{1-x}_0dz\,\Bigg\{\ln\Big[1-x+x\left(M_\xi/m_\psi\right)^2-z(1-x-z)(M_Z/m_\psi)^2\Big]\nonumber\\
&\qquad\qquad\qquad\qquad\qquad-\frac{m_\psi^2+z(1-x-z)M_Z^2}{m_\psi^2\left[1-x+x(M_\xi/m_\psi)^2-z(1-x-z)(M_Z/m_\psi)^2\right]}\Bigg\}\Bigg]\nonumber\\
=&~-\frac{y_\mu^2}{16\pi^2}\Bigg[\mathcal{J}\left(\frac{M^2_\xi}{m_\psi^2}\right)+\int^1_0dx\Big\{\mathcal{R}(x)-\mathcal{N}(x)\Big\}\Bigg]~.
\end{align} 
As before, the condition $Y_\psi=Y_{\mu_R}$ is crucial to remove the UV divergence. Here $\mathcal{R}(x)$ is the same as of Eq.~\eqref{eq:Mx}, and 
\begin{align}
\mathcal{N}(x)=x-1+\left(\frac{2-x\left\{1-(M_\xi/m_\psi)^2\right\}}{z_+-z_-}\right)\times\ln\left[\frac{z_-(x+z_+-1)}{z_+(x+z_--1)}\right]~,
\end{align}
with 
\begin{align}
z_\pm=\left(\frac{1-x}{2}\right)~\pm~\frac{1}{2(M_Z/m_\psi)}\Bigg[(1-x)^2(M_Z/m_\psi)^2-4\Big\{1-x+x(M_\xi/m_\psi)^2\Big\}\Bigg]^{1/2}~.
\end{align}
Note that $\delta \mathcal{W}^\mathbf{R}_{Z\mu\mu}$ alters only the coupling between the right-handed muons and the $Z$ boson. However, the one-loop diagram produced through the $Z^\prime$ exchange can affect both of the left and right-chiral couplings. Therefore, the $Z\mu^+\mu^-$ vertex correction arising from Fig.~\ref{fig:Zll}\,(c) can be cast as,
\begin{align}
\bar{u}(p_2)\delta \mathcal{U}^\nu_{Z\mu\mu}& u(p_1)=-i\,(g^\prime)^2\bar{u}(p_2)\int\frac{d^4k}{(2\pi)^4}\Bigg[ \gamma_\alpha P_{R,\,L}~\frac{(\slashed{p}_2-\slashed{k}+m_\mu)}{(p_2-k)^2-m_\mu^2}~\gamma^\nu P_{R,\,L}~ \frac{(\slashed{p}_1-\slashed{k}+m_\mu)}{(p_1-k)^2-m_\mu^2}\nonumber\\
&\qquad\qquad\qquad\qquad\qquad\qquad\qquad\qquad\times \frac{\gamma_\beta P_{R,\, L}}{k^2-M_{Z^\prime}^2}\left(g^{\alpha \beta}-\frac{k^\alpha k^\beta}{M_{Z^\prime}^2}\right) \Bigg]u(p_1)\nonumber\\
=&-i\,(g^\prime)^2 \bar{u}(p_2)P_{L,\,R}\int\frac{d^4k}{(2\pi)^4}\left[\frac{\gamma_\alpha(\slashed{p}_2-\slashed{k})\gamma^\nu (\slashed{p}_1-\slashed{k})\gamma_\beta\left(g^{\alpha \beta}-\frac{k^\alpha k^\beta}{M_{Z^\prime}^2}\right)}{(k^2-M_{Z^\prime}^2)\{(p_1-k)^2-m_\mu^2\}\{(p_2-k)^2-m_\mu^2\}}\right]u(p_1)~.
\end{align}
With Feynman parametrization, it changes to,
\begin{align}
\bar{u}(p_2)\delta \mathcal{U}^\nu_{Z\mu\mu}u(p_1)=-2i\,(g^\prime)^2 \bar{u}(p_2)P_{L,\,R}\int^1_0dx\int_0^{1-x}dy\int\frac{d^4P}{(2\pi)^4}\Bigg[\frac{\mathbb{N}^\nu(P)}{(P^2-\Delta_4)^3}\Bigg]u(p_1)~,
\label{eq:zmm2}
\end{align}
where, $\Delta_4=M_{Z^\prime}^2\left[x+(1-x)^2R_\mu-y(1-x-y)R_Z\right]$ with $R_\mu$ and $R_Z$ denoting $(m_\mu/M_{Z^\prime})^2$ and $(M_Z/M_{Z^\prime})^2$, respectively. In general, the numerator $\mathbb{N}^\nu(P)$ contains a large number of terms. However, it can be greatly simplified if one assumes $(m_\mu/M_Z)^2\to 0$. Thus, after the momentum integration, Eq.~\eqref{eq:zmm2} reduces to,
\begin{align}
\delta \mathcal{U}^\nu_{Z\mu\mu} &~\equiv\frac{(g^\prime)^2}{8\pi^2}\,\gamma^\nu P_{R,\,L}\int^1_0 dx\int^{1-x}_0 dy\,\Bigg[\frac{R_Z\{x+y(1-x-y)\}\{2+y(1-x-y)R_Z\}}{2\{x+(1-x)^2R_\mu-y(1-x-y)R_Z\}}\nonumber\\
&+3\{x-y(1-x-y)R_Z\}(\tilde{\Delta}_\zeta+\ln\Delta_4)+\Bigg\{1+R_Z\left[\frac{3x}{2}-1+3y(1-x-y)\right]\Bigg\}(\Delta_\zeta-\ln\Delta_4)\Bigg]\nonumber\\
&=\delta \mathcal{V}_{Z\mu\mu}\gamma^\nu P_{R,\,L}~,
\end{align}  
where, $\tilde{\Delta}_\zeta=-\frac{1}{\zeta}+\gamma_E-1-\ln(4\pi)+\mathcal{O}(\zeta)$. Note that the diverging terms in $\tilde{\Delta}_\zeta$ and $\Delta_\zeta$ are oppositely aligned. In case of $\tilde{\Delta}_\zeta$, the negative divergence stems from $\Gamma(1-d/2)$ as $d\to 4$~\cite{Peskin:1995ev}. Therefore,
\begin{align}
\delta \mathcal{V}_{Z\mu\mu}&~=\frac{(g^\prime)^2}{8\pi^2}\Bigg[\int^1_0 dx\int^{1-x}_0 dy\,\Bigg\{\frac{R_Z[x+y(1-x-y)][2+y(1-x-y)R_Z]}{2[x+(1-x)^2R_\mu-y(1-x-y)R_Z]}\nonumber\\
&+\left[3x(1-R_Z/2)+R_Z-1\right]\times\ln\left[x+(1-x)^2R_\mu-y(1-x-y)R_Z\right]\nonumber\\
&-6y(1-x-y)R_Z\times\ln\left[x+(1-x)^2R_\mu-y(1-x-y)R_Z\right]\Bigg\}+\frac{1}{2}\left(\tilde{\Delta}_\zeta+\ln M_{Z^\prime}^2\right)\nonumber\\
&+\frac{1}{2}\left(\Delta_\zeta-\ln M_{Z^\prime}^2\right)+R_Z\Big\{-\frac{1}{4}\left(\Delta_\zeta-\ln M_{Z^\prime}^2\right)+\frac{1}{8}\left(\Delta_\zeta-\tilde{\Delta}_\zeta-2\ln M_{Z^\prime}^2\right)\Big\}\Bigg]\nonumber\\
&~=\frac{(g^\prime)^2}{8\pi^2}\Bigg[\frac{R_Z}{8}+\int^1_0 dx\int^{1-x}_0 dy\Bigg\{\frac{R_Z[x+y(1-x-y)][2+y(1-x-y)R_Z]}{2[x+(1-x)^2R_\mu-y(1-x-y)R_Z]}\nonumber\\
&+\left[3x(1-R_Z/2)+R_Z-1\right]\times\ln\left[x+(1-x)^2R_\mu-y(1-x-y)R_Z\right]\nonumber\\
&-6y(1-x-y)R_Z\times\ln\left[x+(1-x)^2R_\mu-y(1-x-y)R_Z\right]\Bigg\}+\frac{1}{2}\left(\tilde{\Delta}_\zeta+\Delta_\zeta\right)\Bigg]~.
\end{align}
Further, the muon self-energy correction in the presence of $Z^\prime$~[i.e., Fig.~\ref{fig:Zll}\,(a)] can be obtained as 
\begin{align}
-i\,\Sigma^{\mu}_2(\slashed{p})&=-(g^\prime)^2\int\frac{d^4k}{(2\pi)^4}\left[\frac{\gamma_\alpha(\slashed{p}-\slashed{k})\gamma_\beta\left(g^{\alpha \beta}-\frac{k^\alpha k^\beta}{M_{Z^\prime}^2}\right)}{(k^2-M_{Z^\prime}^2)\{(p-k)^2-m_\mu^2\}}\right]\nonumber\\
&=\frac{i(g^\prime)^2}{16\pi^2}\int^1_0dx\Bigg[2x(\Delta_\zeta-\ln\Delta_5)+\frac{1}{M_{Z^\prime}^2}\Big\{x(1-x)^2p^2\,(\Delta_\zeta-\ln\Delta_5)\nonumber\\
&\qquad\qquad\qquad\qquad\qquad\qquad\qquad\qquad+(4-3x)\Delta_5\,(\tilde{\Delta}_\zeta+\ln\Delta_5)\Big\}\Bigg]\slashed{p}~,
\end{align}
where, $\Delta_5=M_{Z^\prime}^2\left[x+(1-x)R_\mu-x(1-x)(p/M_{Z^\prime})^2\right]$. Therefore, following the on-shell renormalization method, one can construct the counter term as, 
\begin{align}
\delta_Z^{\mu(2)}=-\frac{(g^\prime)^2}{8\pi^2}\Bigg[\frac{1}{2}(\Delta_\zeta+\tilde{\Delta}_\zeta)+\frac{1}{2}\int^1_0dx\,\Big\{x(2-3x)\ln\left[x+(1-x)^2R_\mu\right]\Big\}+\mathcal{O}(m_\mu^2)+\cdots\Bigg]\,.
\end{align}
Thus, the renormalized vertex correction factor is given by,
\begin{align}
\delta \mathcal{V}^{\mathbf{R}}_{Z\mu\mu}&=\delta \mathcal{V}_{Z\mu\mu}+\delta_Z^{\mu(2)}\nonumber\\
&=\frac{(g^\prime)^2}{8\pi^2}\Bigg[\frac{R_Z}{8}-\frac{1}{2}\int^1_0dx\,\Big\{x(2-3x)\ln\left[x+(1-x)^2R_\mu\right]\Big\}\nonumber\\
&+\int^1_0 dx\int^{1-x}_0 dy\,\Bigg\{\frac{R_Z[x+y(1-x-y)][2+y(1-x-y)R_Z]}{2[x+(1-x)^2R_\mu-y(1-x-y)R_Z]}\nonumber\\
&+\left[3x(1-R_Z/2)+R_Z-1\right]\times\ln\left[x+(1-x)^2R_\mu-y(1-x-y)R_Z\right]\nonumber\\
&-6y(1-x-y)R_Z\times\ln\left[x+(1-x)^2R_\mu-y(1-x-y)R_Z\right]\Bigg\}\Bigg]\nonumber\\
&=\frac{(g^\prime)^2}{8\pi^2}\Bigg[\frac{R_Z}{8}+\int^1_0 dx\,\Big\{\mathcal{P}_1(x)+\mathcal{P}_2(x)+\mathcal{P}_3(x)+\mathcal{P}_4(x)\Big\}\Bigg]~.
\label{eq:zmmfin}
\end{align} 
Here, the $\mathcal{P}_i$~[$i=1,2,3,4$] functions stand for
\begin{align}
\mathcal{P}_1(x)&=-\frac{x(2-3x)}{2}\,\ln\left[x+(1-x)^2R_\mu\right]\,,\nonumber\\
\mathcal{P}_2(x)&=-\frac{R_Z(1-x)^3}{12}-\frac{(1-x)}{2}\left[2+x(R_Z+1)+(1-x)^2R_\mu\right]\nonumber\\
&+\frac{1}{2(y_+-y_-)}\left[\Bigg\{1+(1-x)^2R_\mu+\frac{(2+R_Z)x}{2}\Bigg\}^2-\left(1-\frac{xR_Z}{2}\right)^2\right]\times\ln\left[\frac{y_-(x+y_+-1)}{y_+(x+y_--1)}\right]\,,\nonumber\\
\mathcal{P}_3(x)&=\left[3x(1-R_Z/2)+R_Z-1\right]\times\Bigg[(1-x)\ln\left\{(1-x-y_+)(1-x-y_-)\right\}-2(1-x)\nonumber\\
&\qquad\qquad\qquad\qquad\qquad\qquad\qquad\qquad+y_+\ln\left(\frac{y_+}{x+y_+-1}\right)+y_-\ln\left(\frac{y_-}{x+y_--1}\right)\Bigg]\,,\nonumber\\
\mathcal{P}_4(x)&=\sum_{k\,=\,+,\,-}\mathcal{E}_k(x)\,,
\label{eq:funcP}
\end{align}
where,
\begin{align}
\mathcal{E}_k(x)&~=2R_Z\left[(1-x-y_k)^3\ln(1-x-y_k)+y_k^3\ln(-y_k)-\frac{1}{3}\left\{(1-x-y_k)^3 +y_k^3\right\}\right]\nonumber\\
&+3R_Z(2y_k+x-1)\left[(1-x-y_k)^2\ln(1-x-y_k)-y_k^2\ln(-y_k)-\frac{(1-x)(1-x-2y_k)}{2}\right]\nonumber\\
&+6R_Z\times y_k(x+y_k-1)\left[(1-x-y_k)\ln(1-x-y_k)+y_k\ln(-y_k)+x-1\right]\,,
\end{align}
and
\begin{align}
y_\pm=\left(\frac{1-x}{2}\right)~\pm~\frac{1}{2\sqrt{R_Z}}\Big[(1-x)^2(R_Z-4R_\mu)-4x\Big]^{1/2}~.
\end{align}
However, the last term of $\mathcal{P}_2(x)$ is discontinuous in the range $x\in[0,\,1]$, resulting in a divergent $x$-integral. Therefore, to ensure numerical stability $\mathcal{P}_2(x)$ has been truncated as,
\begin{align}
\mathcal{P}_2(x)&=-\frac{R_Z(1-x)^3}{12}-\frac{(1-x)}{2}\left[2+x(R_Z+1)+(1-x)^2R_\mu\right]~.
\label{eq:P2}
\end{align} 
It has been numerically checked that for small $R_Z$ values~(i.e., for $M_{Z^\prime}\geq \mathcal{O}(100)$ GeV), there is no significant difference between the results generated with Eq.~\eqref{eq:funcP} and Eq.~\eqref{eq:P2}. However, for higher $R_Z$ values, the numerical instability is largely amplified, leading to physically unacceptable results.

Following Eq.~\eqref{eq:NPe1}, the complete BSM correction to the $Z\to\mu^+\mu^-$ decay can be formulated as,
\begin{align}
\Delta \Gamma_{Z\mu\mu}&=\frac{G_FM_Z^3}{3\sqrt{2}\,\pi}~\Bigg|\left(A_R^2+A_L^2\right)\left(\left|\delta \mathcal{V}^{\mathbf{R}}_{Z\mu\mu}\right|^2+2\,{\rm Re}\left[\delta \mathcal{V}^{\mathbf{R}}_{Z\mu\mu}\right]\right)\nonumber\\
&\qquad\qquad+A_R^2\left(\left|\delta \mathcal{W}^{\mathbf{R}}_{Z\mu\mu}\right|^2+2\,{\rm Re}\left[\delta \mathcal{W}^{\mathbf{R}}_{Z\mu\mu}\right]+2\,{\rm Re}\left[\left(\delta \mathcal{V}^{\mathbf{R}}_{Z\mu\mu}\right)^*\times\delta \mathcal{W}^{\mathbf{R}}_{Z\mu\mu}\right]\right)\Bigg|~.
\end{align}
In general, $\Delta \Gamma_{Z\mu\mu}$ is a function of five independent parameters --- $g^\prime$, $M_{Z^\prime}$, $y_\mu$, $M_\xi$, and $m_\psi$. However, $m_\psi$ is correlated to $M_\eta$ via Eq.~\eqref{cancel2} to ensure a vanishingly small DD cross section. Thus, with $M_\eta=100$ GeV, one has to fix $m_\psi=1.68$ TeV. Moreover, for the $U(1)_{L_\mu-L_\tau}$-specific parameters, one must take care of the existing experimental constraints~[see Fig.~\ref{fig:U1}]. 
\begin{figure}[!ht]
\centering
\subfloat[(a)]{\includegraphics[scale=0.66]{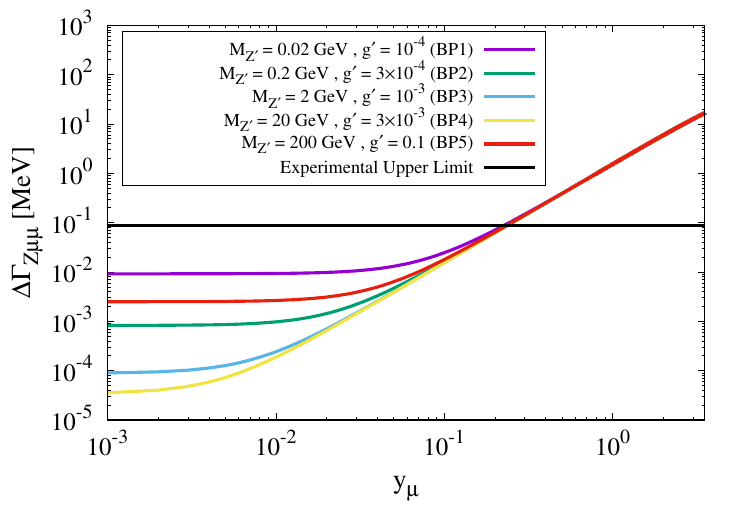}}~~
\subfloat[(b)]{\includegraphics[scale=0.66]{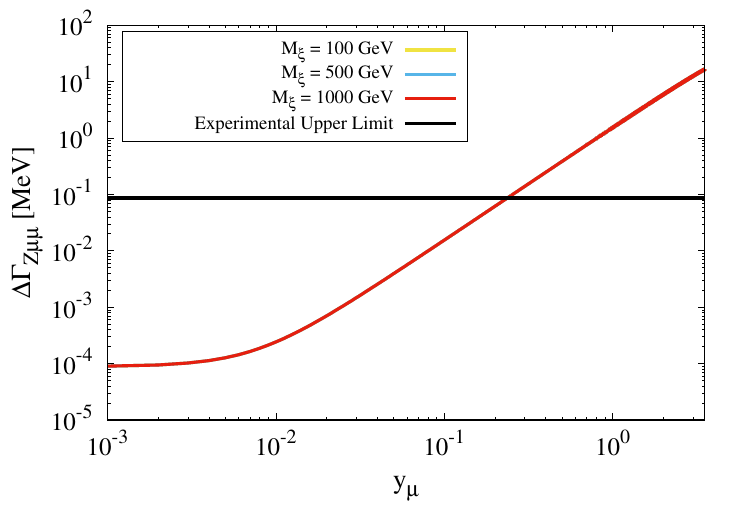}}
\caption{Variation of $\Delta\Gamma_{Z\mu\mu}$ as a function of the NP coupling $y_\mu$ (a) for $M_\xi=150$ GeV with the different colors representing the five sets of $\left(M_{Z^\prime},\,g^\prime\right)$ values and, (b) for $\left(M_{Z^\prime}=2~{\rm GeV},\, g^\prime=10^{-3}\right)$ with $M_\xi=100$ GeV~(yellow), 500 GeV~(sky blue), and 1000 GeV~(red). The horizontal black line at 0.086 MeV shows the experimental upper limit on $\Delta\Gamma_{Z\mu\mu}$. Here, $m_\psi=1.68$ TeV has been considered for the analysis.}
\label{fig:Zmm}
\end{figure}
The variation of $\Delta\Gamma_{Z\mu\mu}$ as $y_\mu$ varies between $\left[10^{-3},\, \sqrt{4\pi}\,\right]$ has been shown in Fig.~\ref{fig:Zmm}. Fig.~\ref{fig:Zmm}\,(a) depicts the variation for $M_\xi=150$ GeV, with violet, green, sky blue, yellow, and red representing $\left(M_{Z^\prime}=0.02~{\rm GeV},\, g^\prime=10^{-4}\right)$, $\left(M_{Z^\prime}=0.2~{\rm GeV},\, g^\prime=3\times 10^{-4}\right)$, $\left(M_{Z^\prime}=2~{\rm GeV},\, g^\prime=10^{-3}\right)$, $\left(M_{Z^\prime}=20~{\rm GeV},\, g^\prime=3\times 10^{-3}\right)$, and $\left(M_{Z^\prime}=200~{\rm GeV},\, g^\prime=0.1\right)$, respectively. For small $y_\mu$ values, $\delta \mathcal{V}^{\mathbf{R}}_{Z\mu\mu}$ acts as the dominant factor in $\Delta\Gamma_{Z\mu\mu}$ with the order of dominance showing a non-trivial dependence on $\left(M_{Z^\prime},\,g^\prime\right)$. However, as $y_\mu$ increases, the NP starts to dominate over the minimal $U(1)_{L_\mu-L_\tau}$ contribution and for $y_\mu\geq 0.3$~(approximately), $\Delta\Gamma_{Z\mu\mu}$ becomes effectively independent of $\left(M_{Z^\prime},\,g^\prime\right)$.    
Fig.~\ref{fig:Zmm}\,(b) shows the case when $\left(M_{Z^\prime},\,g^\prime\right)$ has been fixed at $(2~{\rm GeV},\, 10^{-3})$ with a varying $M_\xi$. However, the plot indicates that the results have no visible dependence on $M_\xi$. As marked by the black line, the experimental bound on the $Z\to \mu^+\mu^-$ decay constrains the NP Yukawa coupling as $y_\mu\leq 0.22$ for $\left(M_{Z^\prime}=0.02~{\rm GeV},\, g^\prime=10^{-4}\right)$. Note that, though the bound can be slightly relaxed for the other $\left(M_{Z^\prime},\,g^\prime\right)$ values, 0.22 will be considered as a general upper limit of $y_\mu$ in all the ensuing computations.
\subsection{$\mathbf{Z\to\tau^+\tau^-}$}
\label{sec:Ztt}
The $\tau$-sector is not affected by the NP Yukawa interactions of Eq.~\eqref{eq:NP} and gets corrected only through the minimal $U(1)_{L_\mu-L_\tau}$ theory. Therefore, for the $Z\to \tau^+\tau^-$ decay, one should only consider Figs.~\ref{fig:Zll}\,(a) and \ref{fig:Zll}\,(c). Thus, the renormalized BSM contribution to $Z\tau^+\tau^-$ vertex, i.e., $\delta \mathcal{V}^{\mathbf{R}}_{Z\tau\tau}$ can be directly replicated from Eq.~\eqref{eq:zmmfin} with a replacement of $R_\mu$ with $R_\tau=\left(m_\tau/M_{Z^\prime}\right)^2$. Therefore, the $U(1)_{L_\mu-L_\tau}$-correction to $\Gamma_{Z \tau \tau}$ can be defined as,
\begin{align}
\Delta \Gamma_{Z\tau\tau}=\frac{G_FM_Z^3}{3\sqrt{2}\,\pi}\left(A_R^2+A_L^2\right)\Bigg|\left|\delta \mathcal{V}^{\mathbf{R}}_{Z\tau\tau}\right|^2+2\,{\rm Re}\left[\delta \mathcal{V}^{\mathbf{R}}_{Z\tau\tau}\right]\Bigg|~.
\label{eq:zttdec}
\end{align}
\begin{figure}[!ht]
\centering
\includegraphics[scale=1]{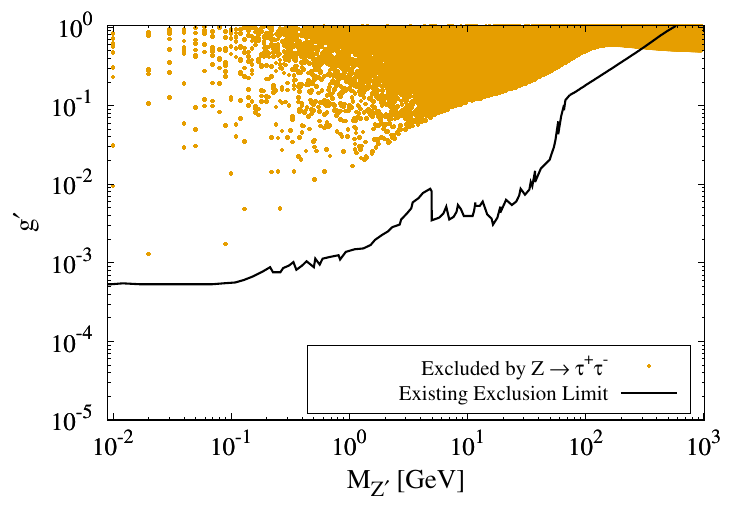}
\caption{Exclusion limit on the $M_{Z^\prime}-g^\prime$ parameter space from the $Z\to \tau^+\tau^-$ decay. The solid black line shows a compilation of the existing experimental bounds~\cite{CCFR:1991lpl,NA64:2024klw, BaBar:2016sci, ATLAS:2023vxg} on the considered plane.}
\label{fig:Ztt}
\end{figure}
As the one-loop correction to the $Z\to \tau^+\tau^-$ decay depends only on the $g^\prime$ and $M_{Z^\prime}$, the experimental bound on $\Gamma_{Z\tau\tau}$ can be used to constrain the minimal $U(1)_{L_\mu-L_\tau}$ model. Fig.~\ref{fig:Ztt} displays the parameter space points~(golden dots) for which $\Delta \Gamma_{Z\tau\tau}\geq 0.086$ MeV. The black line is a compilation of the most stringent exclusion limits from various experimental searches~\cite{CCFR:1991lpl,NA64:2024klw, BaBar:2016sci, ATLAS:2023vxg} for a hidden $U(1)_{L_\mu-L_\tau}$ sector~[see Fig.~\ref{fig:U1}]. Although for the lighter $Z^\prime$ masses, the region excluded by $Z\to \tau^+\tau^-$ decay has already been covered by the existing experiments, for $M_{Z^\prime}\geq \mathcal{O}(100)$ GeV, it supersedes the CCFR bound~\cite{CCFR:1991lpl}. Indeed, it is a notable result as in the parameter space $\{M_{Z^\prime}\geq 300~{\rm GeV},\, g^\prime\geq 0.5\}$, neutrino trident production was known to produce the most stringent exclusion limit. To the best of the author's knowledge, for the first time in the literature, a stronger existing bound has been reported in the aforementioned parameter space. However, the future muon collider searches will be able to probe this region~\cite{Huang:2021nkl}.

Note that one-loop correction to the $Z\to \ell^+\ell^-$ decays is a common characteristic of all the $U(1)_{L_i-L_j}$ extensions of the SM. Therefore, Eq.~\eqref{eq:zttdec} can be easily adapted for $U(1)_{L_e-L_\mu}$ and $U(1)_{L_e-L_\tau}$ models as well to check the consistency of the presently allowed parameter spaces with the bounds on the $Z\to \ell^+\ell^-$ decays. 
\section{Lepton Anomalous Magnetic Moments}
\label{sec:6}
\noindent
The lepton anomalous magnetic moments~$\left(a_\ell\equiv (g-2)_\ell/2\right)$ set a stringent precision test for the SM and can be significant to constrain a BSM theory that affects the SM lepton sector. Out of the three lepton generations, the muon anomalous magnetic moment has been explored to a very high degree of precision, both theoretically and experimentally. Based on the data collected from 2020 to 2023 by the Muon $g-2$ Experiment at the Fermi National Accelerator Laboratory~(FNAL), the latest and most precise experimental world average for $a_\mu$ can be read as, $a_\mu^{\rm Exp}=1165920715(145)\times 10^{-12}$~\cite{Muong-2:2025xyk}. However, the hadronic contributions to $(g-2)_\mu$ being a major source of theoretical uncertainties, there was a long-standing tension between the experimental observation and the SM prediction for $a_\mu$. In the data-driven dispersive evaluations of the leading order hadronic-vacuum-polarization~(LO HVP) contribution, the uncertainties were enhanced through the dominant $e^+e^-\to \pi^+\pi^-$ channel. The problem with the existing data-driven methods was further confirmed through a new measurement of the $e^+e^-\to \pi^+\pi^-$ cross section by the CMD-3 detector~\cite{CMD-3:2023alj}. However, at the same time, the lattice-QCD calculations~\cite{Borsanyi:2020mff,Boccaletti:2024guq} proved to be crucial to reduce the hadronic uncertainty and appeared as a vital computational tool to estimate $a_\mu$. With the substantially increased precision, a consolidated lattice-QCD average of the LO HVP contribution has been attained with a precision of about $0.9\%$. Thus, the current SM prediction for the muon anomalous magnetic moment is given by, $a_\mu^{\rm SM}=116 592 033(62)\times 10^{-11}$~\cite{Aliberti:2025beg}, resulting in a much smaller discrepancy between the SM prediction and the experimental measurements compared to the earlier studies. The current deviation of the $a_\mu^{\rm SM}$ from the experiments can be defined as,
\begin{align}
\Delta a_\mu^{\rm 2025}=a_\mu^{\rm Exp}-a_\mu^{\rm SM}=(3.8\pm 6.3)\times 10^{-10}~.
\end{align}

For the electrons, though the anomalous magnetic moment $a_e$ is experimentally well-measured~\cite{Fan:2022eto}, the SM prediction follows the data-driven methods that rely upon the measurement of the fine-structure constant using the recoil of the atoms~\cite{Aoyama:2019ryr}. Presently, the measurements corresponding to Rubidium-87~\cite{Morel:2020dww} and Cesium-133~\cite{Parker:2018vye} show a $5.5\sigma$ discrepancy between the calculated values of $a_e^{\rm SM}$. The resultant deviations from the experiment can be defined as,
\begin{align}
\Delta a_e^{\rm Cs}&=(-8.8\pm 3.6)\times 10^{-13}~,\nonumber\\
\Delta a_e^{\rm Rb}&=(4.8\pm 3.0)\times 10^{-13}~.
\end{align}
In the present paper, since the electrophilic NP interaction can produce only a negative BSM correction to $a^{\rm SM}_e$, $\Delta a_e^{\rm Cs}$ will be used to constrain the parameter space. 

Unlike the first two generations, measuring $a_\tau$ is extremely challenging due to the short lifetime of $\tau$. The best experimental bound on the $\tau$ anomalous magnetic moment is given by~\cite{DELPHI:2003nah},
\begin{align}
-0.052<a_\tau<0.013~,
\end{align} 
while the corresponding SM prediction is $a_\tau^{\rm SM}=117721(5)\times 10^{-8}$~\cite{Eidelman:2007sb}. For a recent study on the measurement of $a_\tau$, the reader may refer to the Ref.~\cite{Verducci:2023cgx}. Note that the lattice-QCD predictions for the LO HVP contribution to $a^{\rm SM}_e$ and $a^{\rm SM}_\tau$ are also available~\cite{Budapest-Marseille-Wuppertal:2017okr,Giusti:2019hkz,Giusti:2020efo}, but their current precision is not comparable to the data-driven calculations.   
\subsection{Electron}
In the presence of the NP fields $\eta$ and $\chi_2$, BM-$U(1)_{L_\mu-L_\tau}$ can generate a one-loop correction term to $(g-2)_e$. Thus, the observed discrepancy between the SM prediction and the experimental measurements for $a_e$~(to be specific, $\Delta a_e^{\rm Cs}$) can be used to constrain the associated BSM parameters. 
\begin{figure}[!ht]
\centering
\includegraphics[scale=0.62]{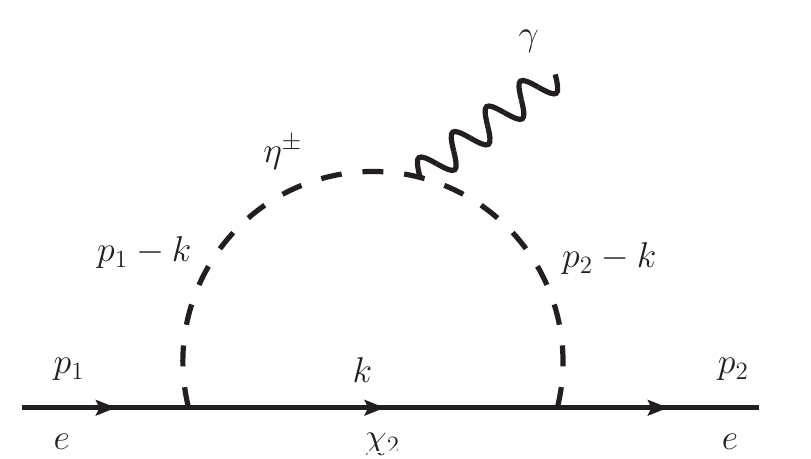}
\caption{One-loop correction to $\gamma e^+e^-$ vertex. $p_1$, $p_2$ denote the external momenta.}
\label{fig:g2e}
\end{figure}
Fig.~\ref{fig:g2e} shows the leading order BSM contribution to $\gamma e^+e^-$ vertex which can be formulated as,
\begin{align}
\Bar{u}(p_2)\,\delta\Gamma^\mu\,& u(p_1)=iy_e^2\,\Bar{u}(p_2)\int\frac{d^4k}{(2\pi)^4}\Bigg[P_L\,\frac{(\slashed{k}+m_2)}{k^2-m_2^2}\,P_R\,\frac{1}{(p_1-k)^2-M_\eta^2} \,\,(p_1+p_2-2k)^\mu\nonumber\\
&\qquad\qquad\qquad\qquad\qquad\qquad\qquad\qquad\qquad\qquad\times\frac{1}{(p_2-k)^2-M_\eta^2}\Bigg] u(p_1),\nonumber\\
&=iy_e^2\,\Bar{u}(p_2)\,P_L\int\frac{d^4k}{(2\pi)^4}\Bigg[\frac{\slashed{k}\,(p_1+p_2-2k)^\mu}{(k^2-m_2^2)\{(p_1-k)^2-M_\eta^2\}\{(p_2-k)^2-M_\eta^2\}}\Bigg]u(p_1)~.
\label{eq:g2e1}
\end{align}
After Feynman parametrization, the NP contribution to the electron anomalous magnetic moment can be obtained as,
\begin{align}
\Delta a_e&~=-i\left(\frac{y_e^2}{2}\right)(2!)\int_0^1\,dx\,(1-x)\int\frac{d^4P}{(2\pi)^4}\Bigg[\frac{2m_e^2\,x(1-x)}{\left(P^2-\Delta_e\right)^3}\Bigg]\nonumber\\
&~=-\frac{y_e^2}{16\pi^2}\left(\frac{m_e}{M_\eta}\right)^2\mathcal{F}(r)~,
\label{eq:a1}
\end{align}
where, $\Delta_e=M_\eta^2[xr+1-x]$ and $r=\left(m_2/M_\eta\right)^2$. The function $\mathcal{F}$ can be defined as, 
\begin{align}
\mathcal{F}(s)=\,\frac{1-6s+3s^2+2s^3-6s^2\ln s}{6(1-s)^4}~.
\end{align}
\begin{figure}[!ht]
\centering
\includegraphics[scale=1]{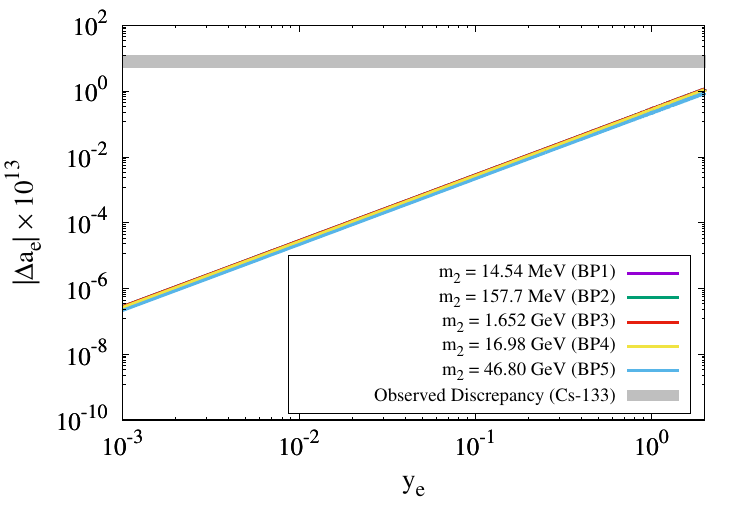}
\caption{Variation of $\Delta a_e$ as a function of $y_e$ for the five different $m_2$ values from Table~\ref{tab:DM} with $M_\eta=100$ GeV. The grey shaded band stands for the $\Delta a_e^{\rm Cs}$.}
\label{fig:g2e_ye}
\end{figure}
Note that the proposed model needs a 3-dimensional parameter space, i.e., $\{M_\eta,\, m_2,\, y_e\}$ to describe the correction to $(g-2)_e$. However, the collider searches have fixed $M_\eta\in[100,\, 107]$ GeV on the lower side while the $Z\to e^+e^-$ decay restricts $y_e$ to be within 2.01. Though in principle, $m_2$ is a free parameter, the benchmark values from Table~\ref{tab:DM} are sufficient to explore the impact of its mass scale on $\Delta a_e$. Fig.~\ref{fig:g2e_ye} depicts the variation of $\Delta a_e$ as a function of $y_e$ for $M_\eta=100$ GeV.
$\Delta a_e$ values corresponding to $m_2=14.54$ MeV, 157.70 MeV, $16.52\times 10^{-1}$ GeV, 16.98 GeV, and 46.80 GeV have been represented with violet, green, red, yellow, and sky blue, respectively. The grey band indicates the observed discrepancy corresponding to Cs-133. Note that in Eq.~\eqref{eq:a1}, a major suppression comes from $(m_e/M_\eta)^2\sim\mathcal{O}(10^{-12})$. Further, with increasing $m_2$, the function $\mathcal{F}$ results in an additional suppression for a given $M_\eta$. Thus, one needs a significantly large $y_e$~$(\geq 5)$ to explain $\Delta a_e^{\rm Cs}$. Therefore, despite generating a BSM contribution to $(g-2)_e$, BM-$U(1)_{L_\mu-L_\tau}$ is not sufficient to address the present observation. 
\subsection{Muon}
\begin{figure}[!ht]
\centering
\subfloat[(a)]{\includegraphics[scale=0.62]{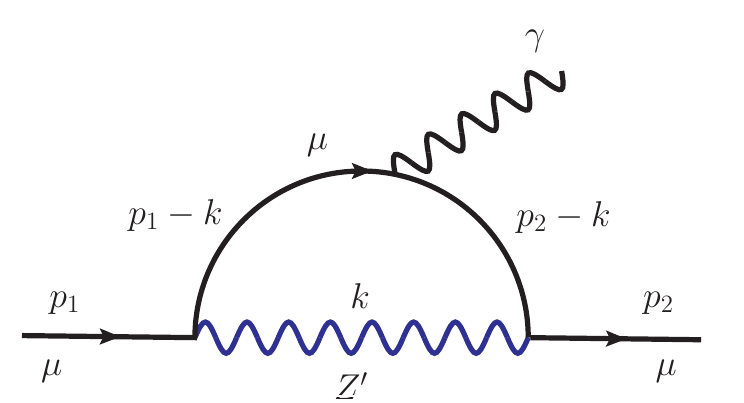}}\qquad\quad
\subfloat[(b)]{\includegraphics[scale=0.62]{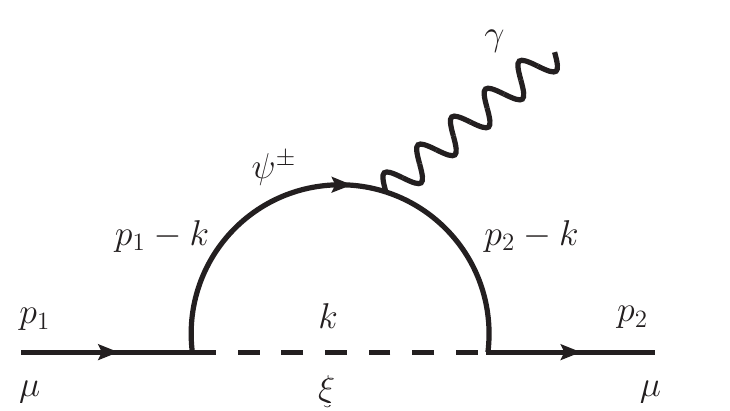}}
\caption{One-loop corrections to $\gamma\mu^+\mu^-$ vertex in the BM-$U(1)_{L_\mu-L_\tau}$ model.}
\label{fig:g2m}
\end{figure}

In the minimal $\mathcal{G}_{\rm SM}\otimes U(1)_{L_\mu-L_\tau}$ model, the only BSM contribution to $\gamma\mu^+\mu^-$ vertex can be obtained at the one-loop level via $Z^\prime$, as shown in Fig.~\ref{fig:g2m}\,(a). 
The corresponding correction to $(g-2)_\mu$ is given by~\cite{Leveille:1977rc, Baek:2001kca},
\begin{align}
\Delta a_\mu^{(1)}=&~\frac{(g^\prime)^2}{8\pi^2}\int_0^1 dx\Bigg[\frac{2m_\mu^2\, x^2(1-x)}{x^2m_\mu^2+(1-x)M_{Z^\prime}^2}\Bigg]\nonumber\\
=&~\frac{(g^\prime)^2\,R_\mu}{4\pi^2}\int_0^1 dx\Bigg[\frac{ x^2(1-x)}{x^2R_\mu+(1-x)}\Bigg]~,
\label{eq:a2}
\end{align}
where, $R_\mu=(m_\mu/M_{Z^\prime})^2$. Fig.~\ref{fig:g2m}\,(b) is the new BSM contribution to $(g-2)_\mu$ that arise within BM-$U(1)_{L_\mu-L_\tau}$. The corresponding vertex correction term can be defined as,
\begin{align}
    \Bar{u}(p_2)\delta\Gamma^\nu u(p_1)&=i\Bar{u}(p_2)\int\frac{d^4k}{(2\pi)^4}\Bigg[\frac{y_\mu P_L\,(\slashed{p}_2-\slashed{k}+m_\psi)}{(p_2-k)^2-m_\psi^2}\,(\gamma^\nu) \,\frac{(\slashed{p}_1-\slashed{k}+m_\psi)}{(p_1-k)^2-m_\psi^2}\times\frac{y_\mu P_R}{k^2-M_\xi^2}\Bigg] u(p_1)\nonumber\\
    &=iy_\mu^2\,\Bar{u}(p_2)\,P_L\int\frac{d^4k}{(2\pi)^4}\Bigg[\frac{(\slashed{p}_2-\slashed{k})\gamma^\nu(\slashed{p}_1-\slashed{k})+m_\psi^2\gamma^\nu}{(k^2-M_\xi^2)\{(p_1-k)^2-m_\psi^2\}\{(p_2-k)^2-m_\psi^2\}}\Bigg]u(p_1)~.
    \label{eq:g2m1}
\end{align}
After Feynman parametrization, the correction to $(g-2)_\mu$ can be read as,
\begin{align}
\Delta a_\mu^{(2)}&~=i\left(\frac{y_\mu^2}{2}\right)(2!)\int_0^1\,dx\,(1-x)\int\frac{d^4P}{(2\pi)^4}\Bigg[\frac{2m_\mu^2\,x(1-x)}{\left(P^2-\Delta_\mu\right)^3}\Bigg]\nonumber\\
&~=\frac{y_\mu^2}{16\pi^2}\left(\frac{m_\mu}{m_\psi}\right)^2\mathcal{F}(w)~,
\label{eq:a3}
\end{align} 
where, $\Delta_\mu=m_\psi^2[xw+1-x]$ and $w=\left(M_\xi/m_\psi\right)^2$.
Therefore, the total BSM contribution to the muon anomalous magnetic moment is given by,
\begin{align}
\Delta a_\mu=\Delta a_\mu^{(1)}+\Delta a_\mu^{(2)}~.
\label{eq:amu}
\end{align}
Note that the same functional structure, $\mathcal{F}$ is appearing in Eqs.~\eqref{eq:a1} and \eqref{eq:a3}. It is a direct consequence of the considered mass hierarchy. If one would have assumed $M_\xi>m_\psi$, the integration in Eq.~\eqref{eq:a3} resulted in a different functional form~\cite{Hisano:1995cp, Kowalska:2017iqv, De:2021crr}. However, it is just a rearrangement of the analytical expression and doesn't alter the numerical results.  
\begin{figure}[!ht]
\centering
\includegraphics[scale=1]{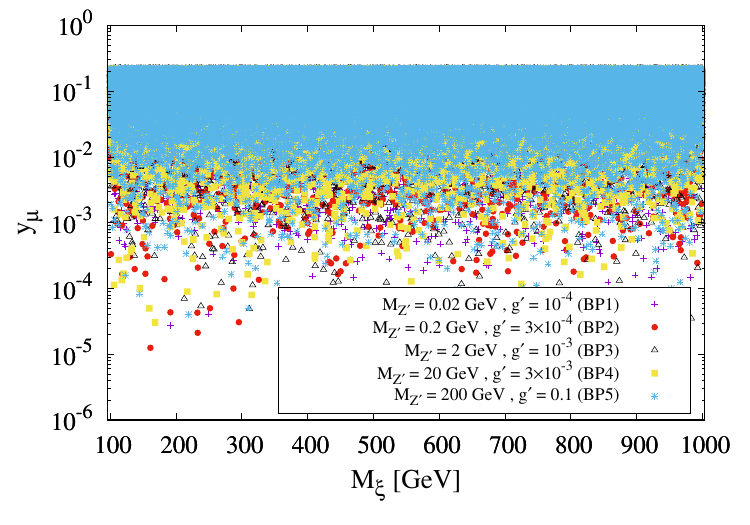}
\caption{$\Delta a_\mu^{\rm 2025}$-satisfying region in the $M_\xi-y_\mu$ plane. The results correspond to $m_\psi=1.68$ TeV.}
\label{fig:g2m_ym}
\end{figure}

The same parameters that govern the BSM contribution to $Z\to\mu^+\mu^-$ also construct the parameter space for $\Delta a_\mu$. Note that the minimal $\mathcal{G}_{\rm SM}\otimes U(1)_{L_\mu-L_\tau}$ model is sufficient to explain the discrepancy in the muon anomalous magnetic moment within the experimentally allowed $\{M_{Z^\prime},\,g^\prime\}$ plane if the recent lattice-QCD results of the LO HVP contribution is considered. However, in the proposed extension of the minimal theory, the additional one-loop contribution to $(g-2)_\mu$ is significant to correlate $\Delta a_\mu$ with an effectively {\it invisible} DM and $\Delta a_e$ through Eq.~\eqref{cancel2}. Fig.~\ref{fig:g2m_ym} displays the parameter space points where $\Delta a_\mu=\Delta a_\mu^{\rm 2025}$. The colors violet, red, black, golden, and sky blue define the $\Delta a_\mu^{(1)}$ contributions corresponding to the five sets of $(M_{Z^\prime},\, g^\prime)$ values from Table~\ref{tab:DM}. Thus, $\Delta a_\mu^{(1)}$ can be associated with a valid DM candidate, while a vanishing DD cross section can be ensured by fixing $m_\psi$ in $\Delta a_\mu^{(2)}$. Therefore, following Eq.~\eqref{cancel2}, $m_\psi$ has been fixed at 1.68 TeV for $M_\eta=100$ GeV. The data points have been generated by varying $y_\mu\in\left[0,\,0.22\right]$ and $M_\xi\in[100,\, 1000]$ GeV.
\begin{figure}[!ht]
\centering
\includegraphics[scale=1]{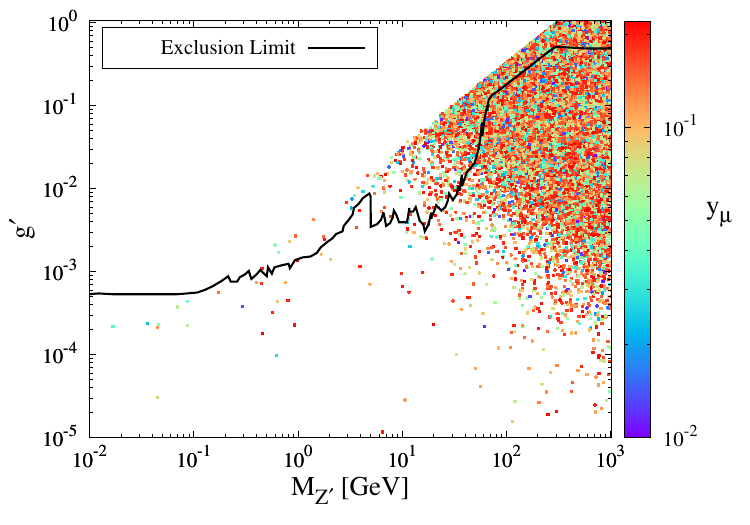}
\caption{The $\Delta a_\mu^{\rm 2025}$-consistent region in a 3-dimensional parameter space. $M_\xi$ and $m_\psi$ have been fixed at 150 GeV and 1.68 TeV, respectively. The solid black line shows the strongest exclusion limit on the $M_{Z^\prime}-g^\prime$ plane --- a compilation of the bounds from the experiments~\cite{CCFR:1991lpl,NA64:2024klw, BaBar:2016sci, ATLAS:2023vxg} and the $Z\to \tau^+\tau^-$ decay.}
\label{fig:g2m_3D}
\end{figure}

Fig.~\ref{fig:g2m_3D} shows a scatter plot in the 3-dimensional parameter space $\{M_{Z^\prime},\, g^\prime,\, y_\mu\}$ with $M_\xi$ and $m_\psi$ being fixed at 150 GeV and 1.68 TeV, respectively. Each point represents a set of parameters for which $\Delta a_\mu=\Delta a_\mu^{\rm 2025}$ while the solid black line marks the excluded region~[see Fig.~\ref{fig:Ztt}]. The $\Delta a_\mu^{\rm 2025}$-satisfying parameter space corresponds to $\left(g^\prime/M_{Z^\prime}\right)^2\leq 1.32\times 10^{-5}$ GeV$^{-2}$ with a subdominant contribution from Eq.~\eqref{eq:a3}. Note that the allowed region is mostly consistent with the exclusion limits for $U(1)_{L_\mu-L_\tau}$ theory. However, the future experimental updates~\cite{Sieber:2021fue,Alekhin:2015byh,Airen:2024iiy,Huang:2021nkl} can be vital to probe a major part of this parameter space and hence, to test the $\mu$-specific NP contribution to $(g-2)_\mu$.
\subsection{Tau}
The considered gauge extension of the $\mathcal{G}_{\rm SM}$ can also produce a BSM contribution to $a_\tau$ at the one-loop level. The Feynman diagram is same as Fig.~\ref{fig:g2m}\,(a) with $\mu$ being replaced by $\tau$. The correction to $a_\tau$ can be cast as,
\begin{align}
\Delta a_\tau=&~\frac{(g^\prime)^2}{8\pi^2}\int_0^1 dx\Bigg[\frac{2m_\tau^2\, x^2(1-x)}{x^2m_\tau^2+(1-x)M_{Z^\prime}^2}\Bigg]\nonumber\\
=&~\frac{(g^\prime)^2\,R_\tau}{4\pi^2}\int_0^1 dx\Bigg[\frac{ x^2(1-x)}{x^2R_\tau+(1-x)}\Bigg]~.
\label{eq:atau}
\end{align}
As before, $R_\tau$ stands for $(m_\tau/M_{Z^\prime})^2$. Note that due to the mass hierarchy among the SM leptons, the $\tau$ anomalous magnetic moment is more sensitive to NP contributions compared to the electron and muon. For BP1~($M_{Z^\prime}=0.02$ GeV \& $g^\prime=10^{-4}$), $\Delta a_\tau=1.22\times 10^{-10}$, while for BP5~($M_{Z^\prime}=200$ GeV \& $g^\prime=0.1$), the value increases by 1 order, resulting in $\Delta a_\tau=6.65\times 10^{-9}$. Therefore, the BSM contribution to $(g-2)_\tau$ is far below the current experimental sensitivity and can't be tested/falsified at present.

\section{Neutrino Masses \& Mixing}
\noindent
\label{sec:neu}
In the presence of the three right-handed Majorana neutrinos with distinct $U(1)_{L_\mu-L_\tau}$ charges~(see Table~\ref{tab:parti}), the NP interactions in the neutrino sector can be defined as~\cite{Biswas:2016yan},
\begin{align}
\mathcal{L}_{\rm N}=&~\frac{1}{2}\Bigg[\sum_{\ell\,=\,e,\,\mu\,\tau}\bar{N}_R^\ell(i\slashed{D})N_R^\ell-m_{ee}(\bar{N}_R^e)^CN_R^e-m_{\mu\tau}\left\{(\bar{N}_R^\mu)^CN_R^\tau+(\bar{N}_R^\tau)^CN_R^\mu\right\}\nonumber\\
&-\lambda_{e\mu}\left\{(\bar{N}_R^e)^C\phi^*N_R^\mu+(\bar{N}_R^\mu)^C\phi^*N_R^e\right\}-\lambda_{e\tau}\left\{(\bar{N}_R^e)^C\phi N_R^\tau+(\bar{N}_R^\tau)^C\phi N_R^e\right\}\Bigg]\nonumber\\
&-\left[\sum_{\ell\,=\,e,\,\mu\,\tau}\beta_\ell~\bar{L}_L^\ell\tilde{H}N_R^\ell+{\rm h.c.}\right]~.
\label{eq:neulag}
\end{align}
Here, $\tilde{H}=i\sigma_2 H^*$, with $\sigma_2$ being the $2^{\rm nd}$ Pauli matrix and the superscript $C$ denotes a charge conjugate state. $\beta_{\ell}$, $\lambda_{e\mu}$, and $\lambda_{e\tau}$ are the dimensionless Yukawa couplings, whereas $m_{ee}$ and $m_{\mu\tau}$ are the dimensionful terms associated with the RHNs. The covariant derivative $D$ can be obtained from Eq.~\eqref{eq:codev}. Note that using the freedom for $\phi$, $Q_\phi$ has been set to $+1$ so that the symmetry breaking scale of $U(1)_{L_\mu-L_\tau}$ can be linked to the neutrino mass and mixing parameters. The choice is consistent with the constraint $Q_\phi\neq \pm|Q_1-Q_2|$. 

The $U(1)_{L_\mu-L_\tau}$ symmetry breaking induces a mixing among the RHNs. The corresponding Majorana mass matrix in the basis $(N_R^e\quad N_R^\mu \quad N_R^\tau)^T$ is given by,
\begin{align}
\mathbf{M}_{\rm RHN}=\left[\begin{array}{c c c}
m_{ee}\qquad & m_{e\mu}\qquad & m_{e\tau}\\
m_{e\mu}\qquad & 0\qquad & m_{\mu\tau}\,e^{i\alpha}\\
m_{e\tau}\qquad & m_{\mu\tau}\,e^{i\alpha}\qquad & 0
\end{array}\right]~,
\end{align}
where, $m_{e\mu}=(\lambda_{e\mu}\,v^\prime)/\sqrt{2}$, and $m_{e\tau}=(\lambda_{e\tau}\,v^\prime)/\sqrt{2}$. Note that in general, all the elements of $\mathbf{M}_{\rm RHN}$ can be complex. However, with a proper phase rotation, one can consider them real with $\alpha$ being the only unremovable phase associated with the $\mu-\tau$ element~\cite{Baek:2015mna}. EWSB mixes the left-handed SM neutrinos and the RHNs through the Yukawa interaction defined in Eq.~\eqref{eq:neulag}. The resulting Dirac mass matrix is diagonal and can be defined as,
\begin{align}
\mathbf{M}_D={\rm diag}\,(\Lambda_e,~\Lambda_\mu,~\Lambda_\tau)~,
\end{align}
where, $\Lambda_\ell=(v\,\beta_\ell)/\sqrt{2}$. Thus, in the 6-dimensional basis of $\left([\nu_\ell]\quad \left[(N_R^\ell)^C\right]\right)^T$, the mass terms of Eq.~\eqref{eq:neulag} can be combined to express as a $6\times 6$ Majorana mass matrix as,
\begin{align}
\mathcal{L}_{\rm N}^{\,\rm mass}=-\frac{1}{2}\left(\left[\bar{\nu}_\ell^C\right]\quad \left[\bar{N}_R^\ell\right]\right)\left[\begin{array}{c c}
\mathbf{0} \qquad & \mathbf{M}_D^T\\
\mathbf{M}_D \qquad & \mathbf{M}_{\rm RHN}
\end{array}\right]\left(\begin{array}{c}
[\nu_\ell]\\
\left[(N_R^\ell)^C\right]
\end{array}\right)+{\rm h.c.}
\label{eq:mass_neu}
\end{align}
Assuming the mass eigenvalues of $\mathbf{M}_{\rm RHN}$ to be much heavier than the EW scale, the mass matrix in Eq.~\eqref{eq:mass_neu} can be block-diagonalized into the light and heavy neutrino mass matrices, defined as
\begin{align}
M_\nu\simeq &~ -\mathbf{M}_D^T\,\left(\mathbf{M}_{\rm RHN}\right)^{-1}\,\mathbf{M}_D\,,\nonumber\\
M_N\simeq &~\mathbf{M}_{\rm RHN}\,,
\end{align}
respectively. 
Using the explicit expressions for $\mathbf{M}_D$ and $\mathbf{M}_{\rm RHN}$, the light neutrino mass matrix in the flavor basis can be obtained as,
\begin{align}
M_\nu=\frac{1}{K}\left[\begin{array}{c c c}
m_{\mu\tau}\Lambda_e^2\,e^{i\alpha} \quad & -m_{e\tau}\Lambda_e\Lambda_\mu \quad & -m_{e\mu}\Lambda_e\Lambda_\tau\\
-m_{e\tau}\Lambda_e\Lambda_\mu \quad & \frac{m_{e\tau}^2}{m_{\mu\tau}}\Lambda_\mu^2\,e^{-i\alpha} \quad & \left(m_{ee}-\frac{m_{e\mu}m_{e\tau}}{m_{\mu\tau}}\,e^{-i\alpha}\right)\Lambda_\mu\Lambda_\tau\\
-m_{e\mu}\Lambda_e\Lambda_\tau \qquad & \left(m_{ee}-\frac{m_{e\mu}m_{e\tau}}{m_{\mu\tau}}\,e^{-i\alpha}\right)\Lambda_\mu\Lambda_\tau \quad & \frac{m_{e\mu}^2}{m_{\mu\tau}}\Lambda_\tau^2\,e^{-i\alpha}
\end{array}\right]~,
\end{align}
where, $K=2m_{e\mu}m_{e\tau}-m_{ee}m_{\mu\tau}\,e^{i\alpha}$. It is a complex symmetric matrix. One has to diagonalize $M_\nu$ to determine the physical light neutrino masses. However, in the considered basis, the correct diagonalization process should involve the Hermitian matrix $\mathbf{S}=M_\nu M_\nu^\dagger$ instead of $M_\nu$~\cite{Xing:2010ez, Adhikary:2013bma}. Therefore, the rotation from the flavor to mass basis can be defined as,
\begin{align}
U^\dagger\,\mathbf{S}\,\,U=\mathcal{D}_\nu={\rm diag}\,(m_{\nu1}^2,~m_{\nu2}^2,~m_{\nu3}^2)\,,
\label{eq:rot}
\end{align}
where, $m_{\nu j}$~($j=1,\,2,\,3$) are the neutrino mass eigenvalues, and
\begin{align}
U=\left[\begin{array}{c c c}
c_{12}c_{13} \quad & s_{12}c_{13} \quad & s_{13}\,e^{-i\delta}\\
-s_{12}c_{23}-c_{12}s_{13}s_{23}\,e^{i\delta} \quad & c_{12}c_{23}-s_{12}s_{13}s_{23}\,e^{i\delta} \quad & c_{13}s_{23}\\
s_{12}s_{23}-c_{12}s_{13}c_{23}\,e^{i\delta} \quad & -c_{12}s_{23}-s_{12}s_{13}c_{23}\,e^{i\delta} \quad & c_{13}c_{23}
\end{array}\right]~.
\end{align} 
Here $s_{ij}=\sin\theta_{ij}$ and $c_{ij}=\cos\theta_{ij}$ with $\theta_{ij}$~($i,\,j=1,\,2,\,3$) defining the three mixing angles. $\delta$ is the Dirac CP violation phase. Note that in principle, there should be the Majorana phases as well, i.e., the rotation defined in Eq.~\eqref{eq:rot} should be through the Pontecorvo-Maki-Nakagawa-Sakata~(PMNS) matrix, $U_{\rm PMNS}=U\times {\rm diag}\,(e^{i\varphi_1},~ e^{i\varphi_2},~ 1)$~\cite{ParticleDataGroup:2024cfk}. $\varphi_{1,\,2}$ are the CP-violating Majorana phases. However, from Eq.~\eqref{eq:rot}, it's trivial to check that $U_{\rm PMNS}\,\mathcal{D}_\nu\,U_{\rm PMNS}^\dagger=U\,\mathcal{D}_\nu\,U^\dagger$. 

In the present framework, neutrino sector can be described by the eight independent model parameters: $\left\{m_{ee},\, m_{\mu\tau},\, m_{e\mu},\, m_{e\tau},\, \Lambda_e,\, \Lambda_\mu,\, \Lambda_\tau,\, \alpha\right\}$. However, there exist seven experimental constraints in terms of the two neutrino mass-squared differences~($\Delta m^2_{ij}=m_{\nu i}^2-m_{\nu j}^2$), three mixing angles, the Dirac CP phase, and the sum of the neutrino masses. Currently, an updated global analysis~\cite{Esteban:2024eli} using the atmospheric neutrino data from Super-Kamiokande and the latest 9.3-year results from IceCube/DeepCore leads to the most stringent bounds on the neutrino masses and mixing angles. Table~\ref{tab:neu} enlists the best fit values in case of normal hierarchy~(NH), i.e., $m_{\nu 3}>m_{\nu 2}>m_{\nu 1}$.    
\begin{table}[!ht]
\centering
\begin{tabular}{|c|c|c|c|c|c|c|}
\hline
Parameters & $\theta_{12}~(^\circ)$ & $\theta_{23}~(^\circ)$ & $\theta_{13}~(^\circ)$ & $\delta~(^\circ)$ & $\Delta m_{21}^2\times 10^{5}~({\rm eV}^2)$ & $\Delta m_{31}^2\times 10^{3}~({\rm eV}^2)$\\
\hline
Best fit $\pm\, 3\sigma$ & $31.63 \to 35.95$ & $41.3 \to 49.9$ & $8.19 \to 8.89$ & $124 \to 364$ & $6.92 \to 8.05$ & $2.451 \to 2.578$\\
\hline
\end{tabular}
\caption{Updated global fit~($3\sigma$ range) for the neutrino mass and mixing parameters in the NH~\cite{Esteban:2024eli}.}
\label{tab:neu}
\end{table}
Further, the cosmological observations result in $\sum_j m_{\nu j}<0.072$ eV~(95\% CL)~\cite{DESI:2024mwx} for the $\Lambda$CDM model. Therefore, assuming $m_{\nu 1}=0$, Eq.~\eqref{eq:rot} can be recast as,
\begin{align}
\left|(\mathbf{S})_{ij}\right|\simeq\left[\begin{array}{c c c}
0.07-0.09 &\qquad 0.21-0.32 &\qquad 0.22-0.27\\
0.21-0.32 &\qquad 1.08-1.49 &\qquad 1.17-1.22\\
0.22-0.27 &\qquad 1.17-1.22 &\qquad 1.08-1.37
\end{array}\right]\times 10^{-3}~.
\label{eq:Smat}
\end{align}
The right-hand side corresponds to the current neutrino constraints~(Table~\ref{tab:neu}) with all the terms being in the ${\rm eV}^2$ unit. Approximately, with $m_{ee}\sim\mathcal{O}(10^4)$ GeV, $m_{\mu\tau}\sim\mathcal{O}(10^2)$ GeV, $m_{e\mu},\,m_{e\tau}\sim\mathcal{O}(10^3)$ GeV, $\Lambda_\ell\sim\mathcal{O}(10^{-4})$ GeV, and $\alpha\in[0,\,2\pi]$, Eq.~\eqref{eq:Smat} can be satisfied for tiny neutrino masses\,\,$\leq\mathcal{O}(10^{-2})$ eV. However, for a detailed numerical analysis, one may refer to Ref.~\cite{Biswas:2016yan}.

\section{Conclusion}
\noindent
\label{sec:7}
The paper has considered $\mathcal{G}_{\rm SM}\otimes U(1)_{L_\mu-L_\tau}$ as the governing gauge theory at the scale of EWSB and extended the particle spectrum with two SM-singlet VLLs $\chi_{1,\,2}$, a charged $SU(2)_L$-singlet VLL $\psi$, an $SU(2)_L$-singlet complex scalar $\eta$, a real SM-singlet scalar $\xi$, and three RHNs $N_R^\ell$. The VLLs and the BSM scalars have been considered odd under an imposed $Z_2$ symmetry, while all the SM fields, RHNs, and a $U(1)_{L_\mu-L_\tau}$-charged SM-singlet scalar $\phi$ are assumed to be even. Moreover, the $\chi_1$ has been kinematically stabilized among the $Z_2$-odd states to be a viable DM candidate. The proposed BSM formulation has been labelled as Beyond the Minimal $U(1)_{L_\mu-L_\tau}$ model, in abbreviation BM-$U(1)_{L_\mu-L_\tau}$. Except $\xi$ and $N_R^e$, all of the NP fields carry non-trivial $U(1)_{L_\mu-L_\tau}$ charges. Further, $\eta$ and $\psi$ being simultaneously charged under both of the abelian gauge groups of the theory, i.e., $U(1)_Y$ and $U(1)_{L_\mu-L_\tau}$, can contribute to the loop-induced kinetic mixing term. The condition for a scale-independent kinetic mixing has correlated the abelian charges of $\eta$ and $\psi$, while their unique determination has followed from the Yukawa interactions of the NP fields with the $1^{\rm st}$ and $2^{\rm nd}$ generation charged leptons. For $m_\psi/M_\eta\to m_\tau/m_\mu$, the kinetic mixing term tends to zero, resulting in an amplitude-level cancellation for the DM-electron and DM-quark scatterings. Thus, the conventional DD methods become invalid in the present framework, and the non-observation of a particle DM in the current and future DD experiments can be trivially explained over the entire parameter space. Note that the DM-muon scattering might be an option to detect $\chi_1$, but the projected sensitivity of the PKU-muon experiment is not adequate to test the present proposal. $\chi_1$ being a $Z^\prime$-portal DM, one can associate the relic density constraint to the $U(1)_{L_\mu-L_\tau}$ gauge parameters. Thus, for the numerical analysis, five benchmark points have been considered from the experimentally allowed region of the $\{M_{Z^\prime},\, g^\prime\}$ parameter space. The corresponding $\chi_1$ masses, for which the observed relic abundance can be satisfied, have been listed, while the $\chi_2$ masses have been fixed through the assumption $m_2=2m_1$. Note that $m_2$ is a free parameter, and in principle, can assume any possible value as long as the considered mass hierarchy between $\chi_1$ and $\chi_2$ is maintained. However, the conclusions remain unchanged for larger $m_2$ values. 
The present model can also generate one-loop correction terms to $(g-2)_\ell$ and $Z\to \ell^+\ell^-$ decay for all three lepton generations. For the electrons, the BSM contributions to $a_e$ and $Z\to e^+e^-$ originate solely from the NP Yukawa coupling, while for the muons, the net BSM corrections are the sum of the one-loop contributions obtained through the $Z^\prime$ exchange and the $\mu$-specific Yukawa interaction. However, the corrections in the $\tau$ sector are the same as those arising in the minimal $U(1)_{L_\mu-L_\tau}$ theory, and hence, can be expressed as a function of $g^\prime$ and $M_{Z^\prime}$ only. In the BM-$U(1)_{L_\mu-L_\tau}$, $Z\to \ell^+\ell^-$ decays are crucial to constrain the NP Yukawa couplings as well as the $\{M_{Z^\prime},\,g^\prime\}$ parameter space. Thus, using the experimental bound on the $Z\to\tau^+\tau^-$ decay, a more stringent exclusion limit has been set on the minimal $U(1)_{L_\mu-L_\tau}$ model. Though the proposed extension is consistent with the recent $(g-2)_\mu$ updates, it results in a subdominant contribution to $(g-2)_e$ for $M_\eta=100$ GeV~(set by the colliders) and $y_e\leq 2.01$. The three RHNs, acquiring masses at an energy scale much higher than that of the EWSB, can couple to the SM neutrinos through a Higgs-mediated Yukawa interaction, whereas the $U(1)_{L_\mu-L_\tau}$ symmetry breaking is vital to induce a mixing among the RHNs. Following the Type-I seesaw mechanism,  the present model can explain the experimental constraints on the neutrino mass and mixing parameters.

Future experiments searching for a $U(1)_{L_\mu-L_\tau}$ sector, improved measurements of $Z\to \ell^+\ell^-$ decays, future updates on the lepton anomalous magnetic moments, and the DD experiments aiming for a $\mu$-philic DM can be significant to test/falsify the BM-$U(1)_{L_\mu-L_\tau}$ model.

\bigskip
\small \bibliography{Darkness}{}
\bibliographystyle{JHEPCust}    
    
\end{document}